\documentclass[letterpaper]{JHEP3}

\usepackage[dvips]{graphicx}
\usepackage{epsfig,multicol,bbm}
\usepackage{amsmath}
\usepackage{appendix}

\def\lsim{\mathrel{\hbox{\rlap{\hbox{\lower4pt\hbox{$\sim$}}}\hbox{$<$}}}}

\makeatletter

\newcommand{\Rmnum}[1]{\expandafter\@slowromancap\romannumeral #1@}
\makeatother

\title{Channeling in direct dark matter detection I: channeling fraction in NaI (Tl) crystals}

\author{Nassim Bozorgnia\\
Department of Physics and Astronomy, UCLA, 475 Portola Plaza, Los
  Angeles, CA 90095, USA\\
E-mail: \email{nassim@physics.ucla.edu}}

\author{Graciela B. Gelmini\\
Department of Physics and Astronomy, UCLA, 475 Portola Plaza, Los
  Angeles, CA 90095, USA\\
Email: \email{gelmini@physics.ucla.edu}}

\author{Paolo Gondolo\\
Department of Physics and Astronomy, University of Utah, 115 South 1400 East \#201,
  Salt Lake City, UT 84112, USA\\
  School of Physics, KIAS, Seoul 130-722, Korea\\
E-mail: \email{paolo@physics.utah.edu}}

\abstract{The channeling of the ion recoiling after a collision with a WIMP changes the ionization signal
in direct detection experiments, producing a larger signal than otherwise expected. We
give estimates of the fraction of channeled recoiling ions in NaI (Tl) crystals
using analytic models produced since  the 1960's and 70's  to describe channeling and blocking effects.
We find that the channeling fraction of recoiling lattice nuclei is smaller than that of ions  that are injected into the  crystal and that it is strongly temperature dependent.}

\begin{document}

\section{Introduction}

Ion channeling in crystals has received a large amount of attention in the interpretation of experiments designed to search for Weakly Interacting Massive dark matter Particles (dark matter WIMPs) through their scattering in a low-background detector. Channeling would occur when the nucleus that recoils after being hit by a dark matter particle moves off in a direction close to a symmetry axis or symmetry plane of the crystal. Channeled ions loose their energy predominantly to electrons, while non-channeled ions transfer their energy to lattice nuclei. In scintillators like NaI (Tl), which are sensitive to the electronic energy losses, channeling increases the fraction of recoil energy that is observed as scintillation light. The presence of channeling in NaI (Tl), for example, affects the mass and cross section of the dark matter particles inferred from the observation of an annual modulation by the DAMA collaboration. As a consequence, the comparison between the DAMA results and the null results of other experiments is affected too~\cite{Savage:2010}.

In addition, ion channeling in crystals could give rise to a daily modulation due to the preferred direction of the dark matter flux arriving on the Earth (this is the direction of the motion of the Solar System with respect to the Galaxy and its dark halo). Earth's daily rotation naturally changes the direction of the ``WIMP wind'' with respect to the crystal axes, thus the amount of recoiling ions that are channeled vs non-channeled, and therefore the amount of energy visible via scintillation. Avignone, Creswick, and Nussinov~\cite{Avignone:2008cw} pointed this out for NaI (Tl) crystals, although the available estimates of the strength of these daily modulations are somewhat simplistic.

Given the importance of channeling in the interpretation of direct detection experiments, and the need to refine the current calculations of a possible daily modulation due to channeling, it is worthwhile to take a deeper look at channeling in the context of dark matter detection. We set out to improve the evaluation of the channeling daily modulation, extend it from NaI (Tl) to Si, Ge and other crystals, and in the process understand channeling in dark matter experiments through analytical means. In this paper, the first one of a series, we present our analytic calculations of the channeling fraction in NaI (Tl). In the subsequent papers, we will present our results for Si, Ge, and other crystals, and our conclusions on the daily modulation due to channeling.

\section{Ion channeling and blocking}

Channeling and blocking effects in crystals refer to the orientation dependence of charged ion penetration in crystals. In the ``channeling effect" ions incident upon a crystal along symmetry axes and planes suffer a series of small-angle scatterings that maintain them in the open ``channels"  between the rows or planes of lattice atoms and thus penetrate much further into the crystal than in other directions. Channeled incoming ions do not get close to lattice sites, where they would be deflected at large angles. The ``blocking effect"  consists in a reduction of the flux of ions originating in lattice sites along symmetry axes and planes, creating what is called a ``blocking dip" in the flux of ions exiting from a thin enough crystal as a function of the exit  angle with respect to a particular symmetry axis or plane.  Directional effects in ion penetration in crystals  was first observed in  1960~\cite{rol} in the sputtering ratio for ions bombarding a single crystal and its explanation in terms of channeling was first done in 1962~\cite{robinson}, although the effect had been predicted to exist in 1912~\cite{stark}. Strongly anisotropic effects for positive particle trajectories originating at lattice sites were discovered in 1965, with particles emitted from radioactive atoms and wide-angle scattering of positive ions in several experiments.  Immediately the relation between these blocking effects and the channeling effect was explained by Lindhard~\cite{Lindhard:1965} in 1965. In the  1960's and 70's  the experimental and theoretical work on channeling proceeded at  a very fast pace (see for example the review by D. Gemmell~\cite{Gemmell:1974ub} and references therein).

  Channeling and blocking effects in crystals are used in crystallography,  in the study  of lattice disorder, ion implantation, and the location of dopant and impurity atoms in crystals, in studies of surfaces, interfaces and epitaxial layers, in measurements of short nuclear lifetimes, in the production of polarized beams etc (see for example~\cite{recent-chann, cohen}).  Channeling and blocking effects are related because the non-channeled incident ions are those which suffer a close-encounter process with an atomic nucleus in the crystal, namely those which pass sufficiently close to a lattice nucleus to be deflected at a large angle. After a close-encounter collision the deflected ion acts as if it was ``emitted" from a lattice site. Channeling is many times observed as a lack of large angle deflections for ions  incident at a small angle $\psi$ with respect to a particular symmetry axis or plane. This forms a ``channeling dip" in the outgoing flux as a function of the incident beam
   angle $\psi$. As first pointed out by Lindhard~\cite{Lindhard:1965}, when no slowing-down processes are involved, the ``channeling" and ``blocking" dips should be identical, when compared for the same particles, energies, crystals and crystal directions.

Ion channeling in NaI (Tl) was first observed in 1973 by Altman, Dietrich, Murray and Rock~\cite{Altman}. They observed that the scintillation output of a monochromatic 10 MeV $^{16}$O beam through an NaI (Tl) scintillator shows two peaks: one at low energy due to non-channeled ions, and one at high energy due to channeled ions. The channeled ions produce more scintillation light because they lose most of their energy via electronic stopping rather than nuclear stopping.

This may be an important effect in  direct dark matter detection experiments in which a scintillation signal due to the recoil of ions as a result of WIMP collisions is searched for. The  potential importance of the channeling effect for direct  dark matter detection was first pointed out for NaI (Tl) by Drobyshevski~\cite{Drobyshevski:2007zj} and by the DAMA collaboration~\cite{Bernabei:2007hw}. When  Na or I ions recoiling after a collision with a  dark matter WIMP move along crystal axes and planes, their quenching factor is approximately $Q=1$ instead of $Q_{\rm Na}=0.3$ and $Q_{\rm I}=0.09$, since they give their energy to electrons. The DAMA collaboration~\cite{Bernabei:2007hw} found that the fraction of channeled recoils is large for low recoiling energies in the keV range, and  that this  effect shifts the regions in cross section versus mass of acceptable WIMP models in agreement with the DAMA data towards lower WIMP masses.

Most of the applications of channeling and blocking are at energies of MeV and higher, however some use much lower energies, up to the keV range. In particular, avoiding channeling is essential in the manufacturing of semiconductor devices, since ion implantation at a controlled depth  is the primary technique. Boron, arsenic and phosphorus ions are implanted in silicon, for example,  to produce integrated circuits, at energies from 100's of eV to several MeV (see for example~\cite{implantation}).

 In this paper we compute  the channeling fraction  of recoiling ions in NaI (Tl)  crystals as function of the recoil energy and the temperature. We do the same for Ge and Si crystals in a companion paper.

 \section{Models of Channeling}

 \subsection{Continuum models}

There are different  approaches to calculate the deflections of ions traveling in a crystal.
 In the ``binary collision model''  the ion path  is computed by a computer program
(see Ref. \cite{Barrett:1971} for one of the first ones) in terms of a succession of  individual interactions, each with one of the atoms in the crystal. Crystal imperfections and lattice vibrations are thus easily and correctly taken into account. In ``continuum models'', reasonable approximations are made which allow to replace the discrete series of binary collisions with atoms by a continuous interaction between a projectile and uniformly charged strings or planes. These models allow to replace the numerical calculations by an analytic description of channeling, and provide good quantitative predictions of the behavior of projectiles in the crystal in terms of  simple physical quantities. This is the approach we use here.  This analytical description was initially developed mostly by J. Lindhard~\cite{Lindhard:1965} and collaborators  for  ions of energy MeV and higher,  and its use was later extended to lower energies, i.e. hundreds of eV and above, mostly to apply it to ion implantation in Si. This approach must be complemented by determination of parameters through data fitting or simulations. Moreover, lattice vibrations are more difficult to include in continuum models. Since we use a continuum model, our results should in last instance be checked by using some of the many sophisticated simulation programs that  implement the binary collision approach or mixed approaches (e.g.~\cite{Monte-Carlo-programs, SARIC, MDRANGE}).

Although the analytical description works better at higher energies  (where it has been very well tested experimentally), at low  and intermediate energies the critical angles for channeling predicted by  analytic models have also been found to be in good agreement  with experimental results.   For the low energy range we
found most useful the work of G. Hobler, who in 1995 and 1996~\cite{Hobler} perfected and checked experimentally previous continuum model
predictions~\cite{cho} for axial and planar channeling at energies in the keV to a few 100 keV range,  to avoid channeling in the implantation of   B, P and As atoms  in Si crystals. Measurements of axial critical angles obtained in the  late 1960's for light (H$^+$,  D$^+$ and He$^+$) and  intermediate mass (B and Ar) ions propagating in various crystals (gold, tungsten,  silicon) with energies between 1 and 100 keV were found to be in good agreement with the predictions of Lindhard models ~\cite{andreen, bergstrom, vanwijngaarden}.   In 1999 K. M. Lui~\cite{lui} and collaborators compared experimental results of 5 keV Ne$^+$ ions  on platinum and predictions of the trajectory simulation code SARIC~\cite{SARIC}, based on the binary collision model, with the predictions of Lindhard's analytical model and the observed angular half-width of the  blocking dips  for axial channels were found to be in good qualitative agreement  with Lindhard's critical angle (both are similar as can be seen in Table I of Ref~\cite{lui}).  In 2002,
S. M. Hogg \textit{et al.}~\cite{hogg} studied channeled implantation of 80 keV Er ions into Si  and concluded that the  axial measured critical angle was in excellent agreement with both computer simulations (made with the MDRANGE program \cite{MDRANGE}) and experimental results. In  2005 Lindhard's critical angle prediction was used to understand qualitative features of  computational results of the SARIC program for  4 keV  Ne$^+$ ions impinging on a Pt surface~\cite{fang}.

Our calculation  is based on  the classical analytic models developed in the 1960's and 70's, in particular by Lindhard~\cite{Lindhard:1965, Andersen:1967, Morgan-VanVliet, VanVliet, Andersen-Feldman, Komaki:1970, Dearnaley:1973, Appleton-Foti:1977}. The fact that the de
Broglie wavelengths of ions in the  keV-MeV range are of the  order of $\sim$ 0.01 pm (and smaller at higher energies), which is much less than the  lattice constant of a crystal ($\sim$ 10  pm), justifies using a classical treatment. We use the continuum string and plane model, in which the
screened Thomas-Fermi potential is averaged over a direction
parallel to a row or a plane. This averaged potential $U$ is considered to be uniformly smeared along the row or plane of atoms, which is a good approximation if the propagating ion interacts with many lattice atoms in the row or plane by a correlated series of many consecutive glancing collisions with lattice atoms. We are going to consider just one row, which simplifies the calculations and is
correct except at the lowest energies we consider, as we explain below.

There are several good analytic approximations of the screened potential. Here we use Lindhard's expression, because it is the simplest and allows to find analytical expressions for the quantities we need. The
  transverse averaged continuum potential of a string as a function of the transverse distance $r$ to the string, relevant for  axial channeling,  was approximated by Lindhard~\cite{Lindhard:1965} as
\begin{equation}
U(r)=E \psi_{1}^2 \, \frac{1}{2}\ln\left(\frac{C^2 a^2}{r^2}+1\right),
\label{axial-pot}
\end{equation}
where $C$ is a constant, which was found experimentally to be $C\simeq\sqrt{3}$, and
\begin{equation}
\psi_{1}^2=\frac{2Z_{1}Z_{2}e^2}{E d}.
\label{psi1}
\end{equation}
$Z_1$, $Z_2$ are the atomic numbers of the recoiling and lattice nuclei respectively, $d$ is the spacing between atoms in the row, $a$ is the Thomas-Fermi screening distance, $a= 0.4685 {\text {\AA} } (Z_1^{1/2} + Z_2^{1/2})^{-2/3} $~\cite{Barrett:1971, Gemmell:1974ub} and   $E= Mv^2/2$ is the kinetic energy of the propagating ion.  In our case, $E$ is the recoil energy imparted to the ion after a collision with a WIMP,
\begin{equation}
E = \frac{|\vec{\bf q}|^2}{2M} ,
\end{equation}
where $\vec{\bf q}$ is the recoil momentum. The string of crystal atoms is at $r=0$.

The transverse averaged continuum potential of a plane of atoms, relevant for planar channeling,  given by Lindhard~\cite{Lindhard:1965} as a function of the distance $x$ perpendicular to the plane is
\label{eq:planar}
\begin{equation}
U_p(x)=E \psi_a^2\left[\left(\frac{x^2}{a^2}+C^2\right)^\frac{1}{2}-\frac{x}{a}\right],
\label{planar-pot}
\end{equation}
where $\psi_a$ is
\begin{equation}
\psi_a=\left(\frac{2\pi n Z_1 Z_2 e^2 a}{E}\right)^\frac{1}{2}
\label{psi_a}
\end{equation}
and $n= N d_{pch}$ is the average number of atoms per unit area, where $N$ is the atomic density and $d_{pch}$ is the width of the planar channel, i.e.  the  interplanar spacing (thus  the average distance of atoms within a plane is $d_p=1/ \sqrt{N d_{pch}}$).
The plane is at $x=0$. Examples of axial and planar continuum potentials are shown in Fig.~\ref{U}.
\FIGURE{\epsfig{file=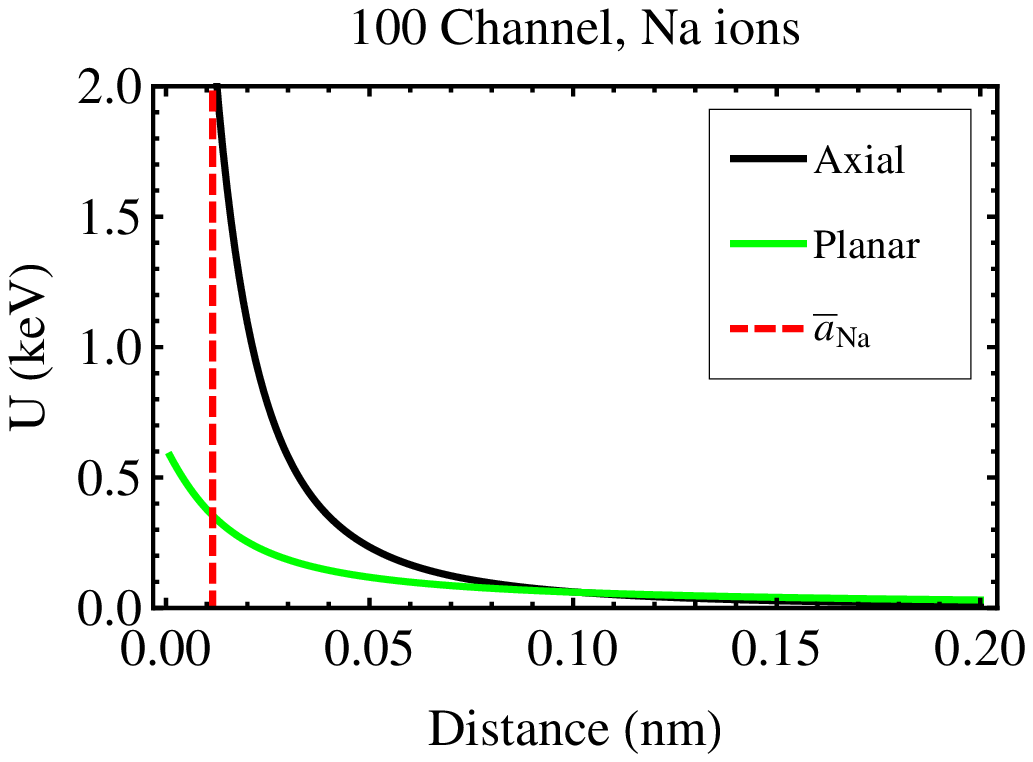,height=139pt}
\epsfig{file=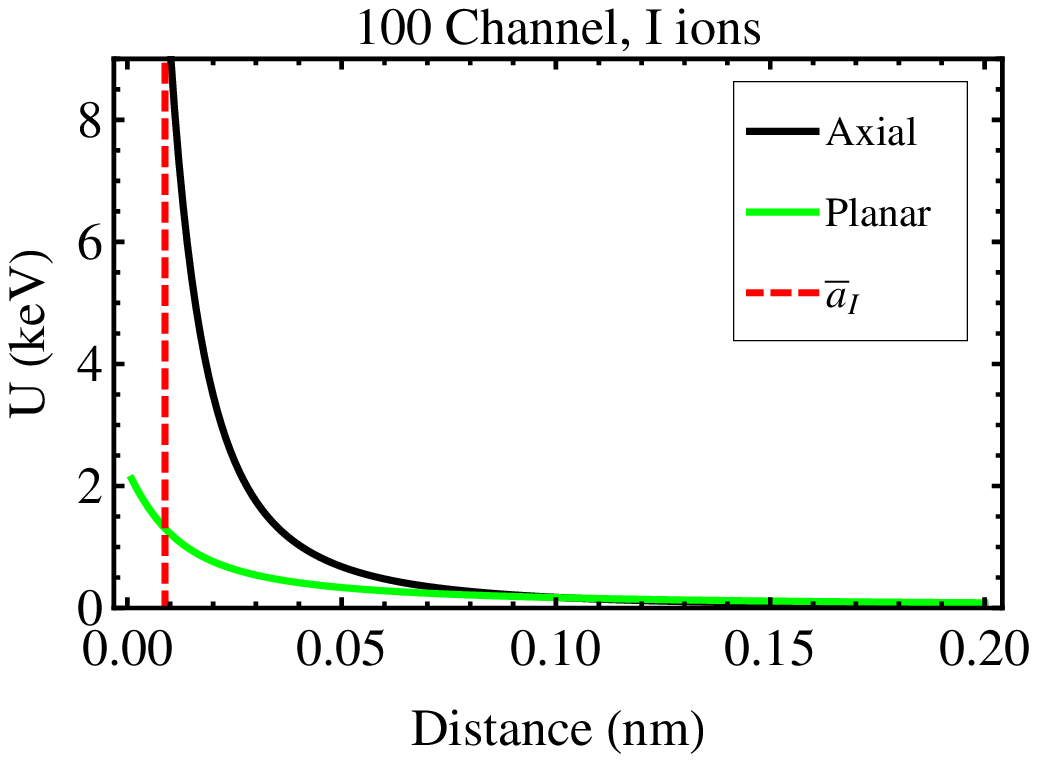,height=135pt}
        \vspace{-0.5cm}\caption{Continuum axial (black) and planar (green/gray) potentials for (a) Na and (b) I ions, propagating  in the $<$100$>$ axial and \{100\} planar channels of an NaI crystal. The screening radii shown as vertical lines are $\bar{a}_{\rm Na}=0.00878$ nm and $\bar{a}_I=0.0115$ nm (see App. A).}
	\label{U}}

 The continuum model  does not imply that the potential energy of an ion moving near an atomic row is well  approximated by the continuum potential $U$. The actual potential consists of sharp peaks  near the atoms and deep valleys in between.  The continuum model says that the net deflection due to the succession of impulses from the peaks is identical to the  deflection due to a force $-U'$. This is only so if the ion never approaches any individual atom so closely that it suffers a large-angle collision.  Lindhard  proved that for a string of atoms this is so only if
 \begin{equation}
U''(r) <  \frac{8}{d^2} E ,
\label{U''}
\end{equation}
where the double prime denotes the second derivative with respect to $r$. Replacing the inequality in Eq.~\ref{U''} by an equality defines an energy dependent critical distance $r_c$
such that  $r >  r_c$ for the continuum model to be valid. Morgan and Van Vliet~\cite{Morgan-VanVliet} also  derived a condition for axial channels, similar to Eq.~\ref{U''} (but with the factor 8 replaced by 16).

The condition in Eq.~\ref{U''} for the validity
of the continuum model on the axial effective potential is equivalent  (as shown initially by Lindhard and proven below) to insuring that the minimum distance of approach to the string remains larger than $\simeq d ~\psi_1$ for large $E$ (MeV and above) and  $\simeq  d ~\psi_2$ (with $\psi_2$ given in Eq.~\ref{psi2}) for small $E$ (below 100's of keV). Thus, the smaller the atomic interdistance $d$ and the larger the ion velocity (i.e. the smaller $\psi_1$ or $\psi_2$) the more accurate the continuum model~\cite{cohen}.

 The appearance of the angle $\psi_1$ in this condition can be easily understood by considering the ``Coulomb shadow" formed by individual atoms behind the direction of arrival of a parallel beam of positive ions. For an unscreened Coulomb potential and small angle deflections, the trajectories of the projectiles give rise to a shadow cone  in which the projectiles do not enter.  A distance $z$ behind the deflecting nucleus in the direction of arrival of the projectiles, the shadow cone radius is $D(z)= \sqrt{2 z d}~ \psi_1$ (here $d$ enters through the definition of $\psi_1$).
As the incident angle of the incoming ions with respect to the row of atoms decreases, there is a critical value of the incident angle at which the edge of the shadow of an atom passes through the adjacent atom. This critical angle is approximately $D(d)/d = \sqrt{2} ~\psi_1$. For angles of incidence larger than $\sim \psi_1$, the shadow cones of the atoms in a row are independent of each other. But for incident angles smaller than $\sim \psi_1$, the shadow cones interfere with each other, so that the atoms in a row are effectively shadowed and not exposed to the projectiles~\cite{ion-book, cohen}. In this case the incident ions do not approach the shadowed
atoms closer than a distance $D(d) \simeq d~\psi_1$, as mentioned above. For a more realistic screened Coulomb potential (as considered in this paper) the shadow cone radius is smaller than for the unscreened potential and the difference between the two becomes smaller for higher ion energies.

The breakdown of the continuum theory for planar channeling is more involved than for axial channeling because  the atoms in the plane contributing to the scattering of the propagating ion are usually displaced laterally within the plane. Thus the moving ion does not encounter atoms at a fixed separation or at fixed impact parameter as is the case for a row.  Morgan and Van Vliet~\cite{Morgan-VanVliet} reduced the problem of scattering from a plane of atoms to the scattering from an equivalent row of atoms contained in a strip centered on the projection of the ion path on the plane of atoms. They then  applied Eq.~\ref{U''}  to the fictitious string defined in this way as the condition for planar channeling (more about this below).
\subsection{The transverse energy}

Lindhard proved that for channeled particles the longitudinal component $v \cos\phi$, i.e. the component along the direction of the row or plane of the velocity, may be treated as constant (if energy loss processes are neglected). Then, in the continuum model, the trajectory of the ions can be completely described in terms of the transverse direction, perpendicular to the row or plane considered. For small  angle $\phi$  between the ion's trajectory and  the atomic  row (or plane)
in the direction perpendicular to  the row (or plane), the  so called ``transverse energy"
\begin{equation}
E_{\perp} = E \sin^2\phi + U\simeq E \phi^2 + U
\label{E-perp}
\end{equation}
is conserved. In Eq.~\ref{E-perp} relativistic corrections are neglected.

In  each binary collision of the ion with the closest atom, $E_{\perp}$ changes
abruptly, because the angle $\phi$ changes in a very short time (i.e. while the potential is practically constant). Then, between two collisions, the change is compensated because the potential component of $E_{\perp}$ changes continuously as the ion propagates while the angle $\phi$ is constant~\cite{cohen}. A good way to test to what extent this compensation takes place is to calculate the value of $E_{\perp}$  far away from the collision sites, namely half-way between successive collision sites.

The condition in Eq.~\ref{U''} was derived by Lindhard (in appendix A of Ref.~\cite{Lindhard:1965}) for axial channels by defining $E_{\perp}$ at the planes half-way between string of atoms
and asking for $E_{\perp}$ at contiguous half-way planes to be conserved to first order (this is  the so called ``half-way plane" model).   Morgan and Van Vliet~\cite{Morgan-VanVliet} derived a condition for axial channeling very similar to Eq.~\ref{U''} by calculating the difference between  the scattering angle due to a binary collision and the deflection angle in the continuum potential when the ion travels the distance between two contiguous halfway planes (the half-way planes considered by Lindhard).

 Let  $r_i$ be the initial position at which the WIMP nucleus collision occurs, i.e. if $r_i >0$ the recoiling nucleus was displaced with respect to its position of equilibrium in the string when it collided with a WIMP. We call $\phi$ the angle of the initial recoil momentum with respect to the row of atoms, and $E$  the initial recoil energy  of the propagating ion.  Given  these initial parameters, the issue of  where to define $E_{\perp}$ arises. Namely, we define
\begin{equation}
E_{\perp}= E \sin^2 \phi + U(r^*),
\label{E-perp-HP}
\end{equation}
but there are different possible choices for $r^*$, the position at which to measure the potential $U$. In the ``half-way plane model" used by Lindhard, $U$ is measured
  after the recoiling  ion propagates a distance $d/2$ along the string, when it is at a distance
\begin{equation}
r^* = r^*_{\rm HP} \equiv r_i + (d/2) \tan{\phi}
\end{equation}
perpendicular to the string at the halfway-plane. All angles we are dealing with are small enough that $\sin\phi \simeq \tan\phi \simeq \phi$. This choice was shown to work better in some respects \cite{Andersen-Feldman} (such as the blocking angular distribution in axial channels)  than the ``continuum approximation." In the latter, the transverse energy $E_{\perp}$ is considered to be conserved all along the string, not only at the halfway-planes, in which case $r^*$ is chosen to be just
\begin{equation}
r^* = r^*_{\rm CA} \equiv r_i.
\end{equation}
The two choices $r^*=r^*_{\rm HP}$ and $r^*=r^*_{\rm CA}$ coincide only if $d \tan\phi/ 2 r_i \ll 1$, a condition that
 at energies below 100's of keV is in general  not fulfilled, in which case the ``continuum approximation"  is not a good approximation. In fact, assuming the ``continuum approximation", the angle at the first halfway plane must have a different value, $\phi'$ say, such that $\sin^2\phi' -\sin^2\phi =  [U(r_i) - U(r_i + \tfrac{1}{2} d \tan{\phi})]/E$. Fig.~\ref{U}  shows that the  potential $U$ at a distance $a$ or larger  is in the 1-10 keV range. Thus the difference between both definitions of the transverse energy is very small at large enough values of the energy $E \gg$ 10 keV, but for lower values of $E$, in the keV to the 10's of keV range,  the definitions $r^*=r^*_{\rm HP}$ and $r^*=r^*_{\rm CA}$ give different results unless  $\phi$ is small enough. In Ref.~\cite{Andersen-Feldman}  the predictions of  both models  for blocking  of 400 keV protons in W and 7 MeV protons in Si were compared with the predictions of the binary-collision model.   The ``half-way plane model" results were found to be in agreement with those of the  binary-collision model, even when those of the  ``continuum approximation" were not.

In all these cases, even when considering blocking, the propagating ion was always different than a lattice ion. In our case, the recoiling ion leaves an empty lattice site, thus it moves away from an empty lattice site in the potential generated by its neighboring lattice atoms. So the potential the recoiling ion moves through at the moment of collision is very small, and the recoiling ion conserves its momentum and direction of motion until it gets very near the nearest neighbor, a distance $d$ away along the string. At this moment, it is at a distance
\begin{equation}
r^* = r^*_{\rm rec}\equiv r_i + d \tan{\phi_i}
\label{r*definition}
\end{equation}
from its nearest neighbor.
Therefore, we will  make the approximation of defining the potential entering into Eq.~\ref{E-perp-HP} at $r^*=r^*_{\rm rec}$.

\subsection{Minimum distance of approach and critical channeling angle}

The conservation of the transverse energy provides a definition of the minimum distance of approach to the string, $r_{\rm min}$ (or to the plane of atoms $x_{\rm min}$), at which the trajectory of the ion makes  a zero angle with the string (or plane), and also of the angle $\psi$ at which the ion exits from the string (or plane), i.e. far away from it where $U \simeq0$. In reality the furthest position from   a string or plane of atoms is  the middle of the channel, whose width we call  $d_{\rm ach}$ for an axial channel or $d_{\rm pch}$ for a planar channel,  thus
\begin{equation}
E_{\perp}= U(r_{\rm min}) =  E \psi^2 +U(d_{\rm ach}/2).
\label{eq:consetrans}
\end{equation}
We define the axial channel width $d_{\rm ach}$ in terms of the interatomic distance $d$ as $d_{\rm ach}= 1/ \sqrt{N d}$, where $N$ is the atomic density.

For axial channeling Lindhard equates the  condition for channeling with  the condition in  Eq.~\ref{U''} for the validity of the continuum model. For Lindhard's axial potential, this condition reads
\begin{equation}
\label{eq:cond}
E > \frac{ E_1 d^2}{8} \, \frac{1+ 3(\frac{r_{\rm min}}{Ca})^2}{r_{\rm min}^2\left(1 + (\frac{r_{\rm min}}{Ca})^2\right)^2} ,
\end{equation}
where $E_1 = E \psi_1^2$ (and  $\psi_1$  was defined in Eq.~\ref{psi1}).
Since the right-hand side of this inequality is a monotonically decreasing function of $r_{\rm min}$, one just needs to solve the equation obtained by replacing the inequality with an equality. Solving a cubic equation, the condition  in Eq.~\ref{eq:cond} can be inverted  to find  $r_c(E)$,   the minimum value of $r_{\rm min}$. We find the following decreasing function of $E$
\begin{equation}
r_c(E) =Ca\, \sqrt{\frac{2}{3} \left[ \sqrt{1+z} \, \cos\!\left( \frac{1}{3} \arccos\frac{(1-3z/2)}{(1+z)^{3/2}} \right) - 1\right] },
\label{rcrit}
\end{equation}
where to simplify the expression we defined
\begin{equation}
z=\frac{9 E_1 d^2}{8 E C^2  a^2} .
\end{equation}

This expression gives the smallest possible minimum distance of approach of the propagating ion with the row for a given energy $E$, i.e. $r_{\rm min} > r_c(E)$
 and, since the potential $U(r)$  decreases monotonically with increasing $r$,
\begin{equation}
U(r_{\rm min}) < U(r_c(E)).
\label{defrcrit}
\end{equation}
Using Eq.~\ref{eq:consetrans}, this can be further translated into an upper bound on $E_{\perp}$ and  on  $\psi$, the angle the ion makes with the string far away from it,
\begin{equation}
\psi < \psi_{c}(E)=  \sqrt{ \frac{U(r_c(E))- U(d_{\rm ach}/2)}{E} }.
\label{psicritaxial}
\end{equation}
 $\psi_{c}(E)$  is the maximum angle  the ion can make with the string far  away from it (i.e. in the middle of the channel) if the ion is channeled.
When $U(d_{\rm ach}/2)$ can be neglected, i.e. when $r_c(E) < d_{\rm ach}/2$, the limiting values of   $\psi_{c}(E)$  (as already proven by Lindhard~\cite{Lindhard:1965}) are   $\psi_{c}(E) \simeq \psi_1$
(see Eq.~\ref{psi1})  for large $E$ ($z\ll 1$, typically close to MeV and larger) and  $\psi_{c}(E) \simeq \psi_2$ at low $E$ ($z\gg 1$, typically smaller than a few  100 keV), where
\begin{equation}
\psi_2 = \sqrt{ \frac{Ca \psi_1}{d \sqrt{2}}}.
\label{psi2}
\end{equation}
One can easily see that the critical  distance $r_c$ becomes $r_c \simeq d~ \psi_1/2 \sqrt{2}$ for large $E$ and  $r_c\simeq  d ~\psi_2$ for small $E$.

The critical distance  $r_c(E)$ increases as $E$  decreases. At low enough $E$,  $r_c(E)$ becomes close to $d_{\rm ach}/2$, and  the critical angle $\psi_{c}(E)$, the maximum angle for channeling in the middle of the channel, goes to zero. This means that there is a minimum energy below which channeling cannot happen, even for ions moving initially in the middle of the channel. This is a reflection of the fact that the range of
the interaction between ion and lattice atoms increases with decreasing energy and at some point there is no position in the crystal where the ion would not be deflected at large angles. The existence of a minimum energy for channeling was found by  Rozhkov and DyulÕdya~\cite{Rozhkov} in 1984 and later by Hobler~\cite{Hobler} in 1996. It is clear that to compute  $r_c(E)$ when it is not small with respect to  $d_{\rm ach}/2$, and thus to compute the actual minimum energy for channeling, we would need to consider the effect of more than one row or plane (as done in Refs.~\cite{Rozhkov} and \cite{Hobler}), thus our results are approximate in this case.

For planar channeling we will follow the definition of fictitious row in Morgan and Van Vliet~\cite{Morgan-VanVliet, Hobler}.
They reduced the problem of scattering from a plane of atoms to the scattering of an equivalent row of atoms contained in a strip of width 2$R$ ($R$ is defined below) centered on the projection of the ion path on the plane of atoms, and took the average area per atom  in the plane to be  2$R$ times the characteristic distance $\bar{d}$ between
atoms along this fictitious row,
\begin{equation}
\bar{d} = 1/ (N d_{\rm pch} 2 R).
\label{dbar}
\end{equation}
 Once the width $2R$ of the fictitious row is specified, one uses the channeling condition for the continuum string model, Eq.~\ref{U''}, with an average atomic composition of the plane. For $R$, Morgan and Van Vliet used the impact parameter in an ion-atom collision corresponding to a deflection of the order of the break-through angle $\sqrt{U_p(0)/E}$. This is the deflection necessary for an ion of energy $E$ approaching the plane from far away (so that the initial potential can be neglected) to overcome the potential barrier at the center of the plane at $x=0$ (namely so that $E_\perp= U_p(0)$). Using the Moliere approximation for the screened potential (which we do not use), Morgan and Van Vliet found for $\bar{d}$
\begin{equation}
\bar{d}^{MV}= \left[A~a N d_{\rm pch} \ln\left(B~Z_1 Z_2 e^2/ a \sqrt{E U_p(0)}\right)\right]^{-1},
\label{dbar-MV}
\end{equation}
with coefficients $A=1.2$ and $B=4$. However, Morgan and Van Vliet~\cite{Morgan-VanVliet} found discrepancies with this theoretical formula  in simulations of binary collisions  of 20 keV protons in a copper crystal and adjusted the coefficients to  $A= 3.6$ and $B=2.5$. Hobler~\cite{Hobler} used both sets of coefficients and compared them with simulations and data of B and P in Si for energies  of about 1 keV and above. Hobler concluded that the original theoretical formula was better in his case. In any case, Hobler proposed yet another empirical relation to define $\bar{d}$.

 In the absence of simulations for NaI, we are going to use an upper bound on $R$, given by the average interdistance of atoms in the plane, $2R < d_p= 1/ \sqrt{N d_{\rm pch}}$, so that replacing the maximum value of $R$ in Eq.~\ref{dbar} we find that the minimum value of $\bar{d}$ is the average  interdistance of atoms in the plane, $d_p$
\begin{equation}
\bar{d}_{\rm min} = d_p.
\label{ourdbar}
\end{equation}

 Thus, for planar channeling,  we use the condition in Eq.~\ref{U''} for a fictitious string, replacing the distance $d$ by the distance $d_p$  and replacing the composition of the string for the average composition of the plane. Let us call $\bar{r}_c(E)$ the critical distance obtained from  Eq.~\ref{rcrit} for this fictitious string, then the minimum distance of approach for planar channeling is
\begin{equation}
x_c(E) \equiv \bar{r}_c(E).
\label{ourxcrit}
\end{equation}
The use  of $\bar{d} = d_p$ yields a lower bound on $x_c$, as shown in Fig.~\ref{rc-MV-ours} (and thus an upper bound on the fraction of channeled recoils as explained later).
  Fig.~\ref{rc-MV-ours} shows the planar critical distances of approach $x_c$ (Eq.~\ref{ourxcrit}) using  the theoretical (with coefficients $A=1.2$ and $B=4$) and adjusted (with coefficients $A= 3.6$ and $B=2.5$) Morgan-Van Vliet expressions for
  $\bar{d}$ in Eq.~\ref{dbar-MV}. Our choice of $x_c$, with $\bar{d}$ in Eq.~\ref{ourdbar} is also plotted in Fig.~\ref{rc-MV-ours}. We can see that it is lower than the others.

\FIGURE{\epsfig{file=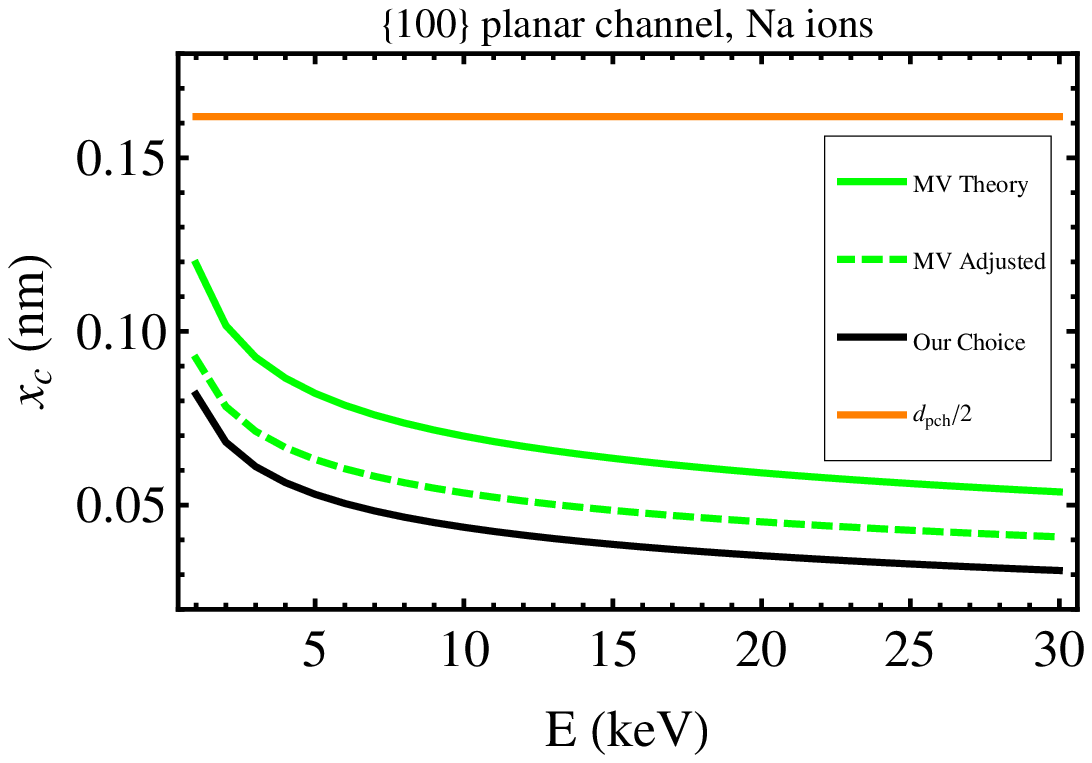,height=210pt}
        \caption{Comparison of planar critical distances of approach as given by the theoretical  (solid line) and adjusted (dashed line) Morgan-Van Vliet (label MV, green/gray) expressions for $\bar{d}$ and our choice (black) of $x_c$, which we take as a lower bound, as function of the energy for Na ions travelling in \{100\} planar channels. Also shown is the radius of the channel $d_{\rm pch}/2$.}
	\label{rc-MV-ours}}

Writing equations equivalent to Eq.~\ref{eq:consetrans} and \ref{defrcrit} for planar channels, namely
\begin{equation}
E_{\perp}= U(x_{\rm min}) =  E (\psi^p)^2 +U_p(d_{\rm pch}/2),
\label{planarEperp}
\end{equation}
and
\begin{equation}
U_p(x_{\rm min}) < U_p(x_c(E)),
\label{planarmincond}
\end{equation}
we obtain an equation similar to Eq.~\ref{psicritaxial} but for the maximum planar channeling angle,
\begin{equation}
\psi^p_c(E)=  \sqrt{ \frac{U_p(x_c(E))- U_p(d_{\rm pch}/2)}{E} }.
\label{psicritplanar}
\end{equation}
For very small energies, for which $x_c(E) \geq d_{\rm pch}/2$ no channeling is possible (the maximum distance  to any plane cannot be larger than half the width of the channel separating them). When $x_c(E)$
 approaches the middle of the channel the effect of other planes should be considered, so our approximation of using the potential of only one plane is not correct in this regime.

 The static lattice critical distances are presented in left panel of  Figs. \ref{rc-psic-MV-100} and  \ref{rc-psic-MV-111} for the
 100 and 111 axial and planar channels.

 There is an alternative way of treating planar channels presented by Matyukhin~\cite{Matyukhin} in 2008, but we have doubts about the validity of this method, for which we have not found any comparison with either simulations or data. For completeness we present it in Appendix C. It predicts larger channeling fractions.

\subsection{Temperature dependent critical distances and angles}

So far we have been considering static strings and planes, but the atoms in a crystal are actually vibrating.  We use here the Debye model to take into account the zero point energy and thermal vibrations of the atoms in a crystal.  The one dimensional rms vibration amplitude $u_1$ of the atoms in a crystal in this model  is~\cite{Gemmell:1974ub, Appleton-Foti:1977}
\begin{equation}
u_1(T)=12.1 \, \text{\AA} \, \left[\left(\frac{\Phi(\Theta/T)}{{\Theta/T}} + \frac{1}{4}\right)(M\Theta)^{-1}\right]^{1/2},
\label{vibu1}
\end{equation}
where the 1/4 term accounts for the zero point energy, $M$ is the atomic mass in amu, $\Theta$ and $T$ are the Debye temperature and the temperature of the crystal in K, respectively, and $\Phi(x)$ is the Debye function,
\begin{equation}
\Phi(x)=\frac{1}{x}\int_{0}^{x}{\frac{t dt}{e^t -1}}.
\end{equation}
Eq.~\ref{vibu1} was derived for monoatomic cubic crystals for which $M$ is clearly specified. In the case of crystals composed of more than one kind of atom, experiments have shown that the difference  of vibration amplitudes of both types is very small for $T>\Theta$~\cite{Lonsdale-1948}, even when the difference of atomic weights of the various  kinds of atoms is large, as is the case for NaI. Using as $M$ the average mass
\begin{equation}
M= (M_{\rm Na} + M_{\rm I})/2
\label{M-average}
\end{equation}
produces an error of less than 10\% in the actual vibration amplitudes at $T > \Theta$~\cite{Lonsdale-1948}. For NaI,  we take the Debye temperature to be $\Theta=165\;$K~\cite{Gemmell:1974ub, Sharko-Botaki-1971} (although it changes with $T$ between 169 K at a few K and 155 K at 300 K~\cite{Sharko-BotakiVar-1971}). The crystals in the DAMA experiment  are at 20 $^\circ$C, i.e. $T=293.15\;$K; $M_{\rm Na}=22.9$ amu and $M_{\rm I}=126.9$ amu, thus $M=74.9$ amu. The vibration amplitude $u_1$ we get using this value of $M$  is plotted in Fig.~\ref{figu1} as a function of the temperature $T$. At room temperature (20 $^\circ$C) it is $u_1=0.0146$ nm which is similar to the measured value of $\sqrt{\langle u_1^2\rangle}=$ 0.0145 nm~\cite{Neelakanda-2008}, while measured separate values of  $\sqrt{ \langle u_1^2\rangle}$  for Na and I  (always at room temperature) are 0.018 nm and 0.015 nm~\cite{Geeta-1998} respectively. (To use the data in Ref.~\cite{Geeta-1998} we must take into account that the  Debye-Waller factor $B$ is  $B= 8 \pi^2 \langle u_1^2\rangle$).
\FIGURE{\epsfig{file=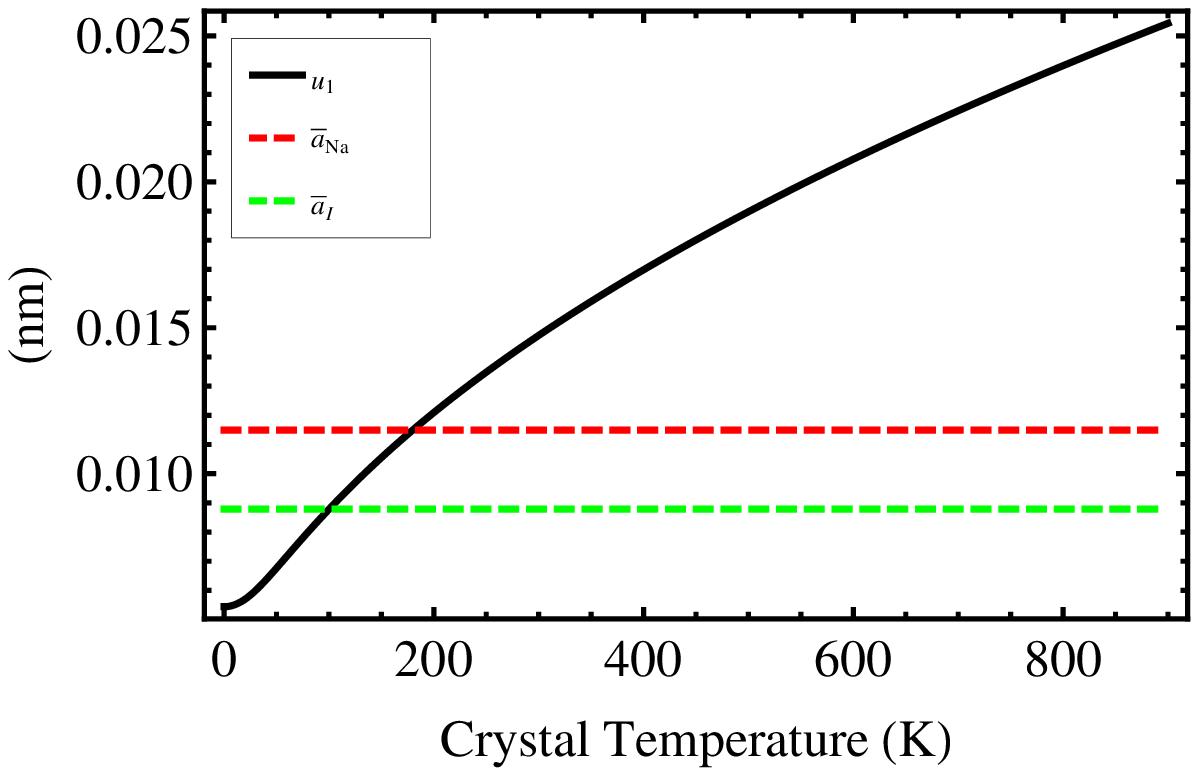,height=210pt}
        \caption[Example of figure]{Plot of $u_1(T)$ for NaI (Eq.~\ref{vibu1} with $M= (M_{\rm Na} + M_I)/2$).}%
	\label{figu1}}

At low temperatures ($T\ll \Theta$)  the individual vibration amplitudes become progressively different~\cite{Hewat-1972}. We are not taking this difference into account in our approach. However, at these temperatures the channeling fractions become so small (as we show below) that a better calculation is not important.

\FIGURE{\epsfig{file=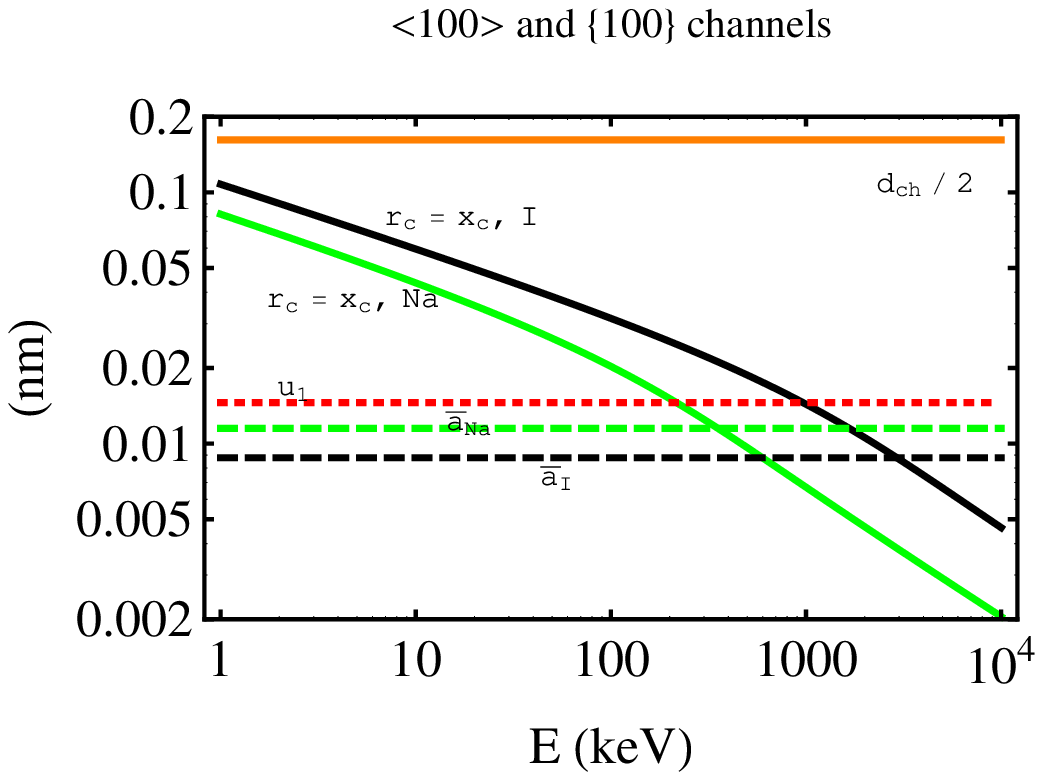,height=160pt}
\epsfig{file=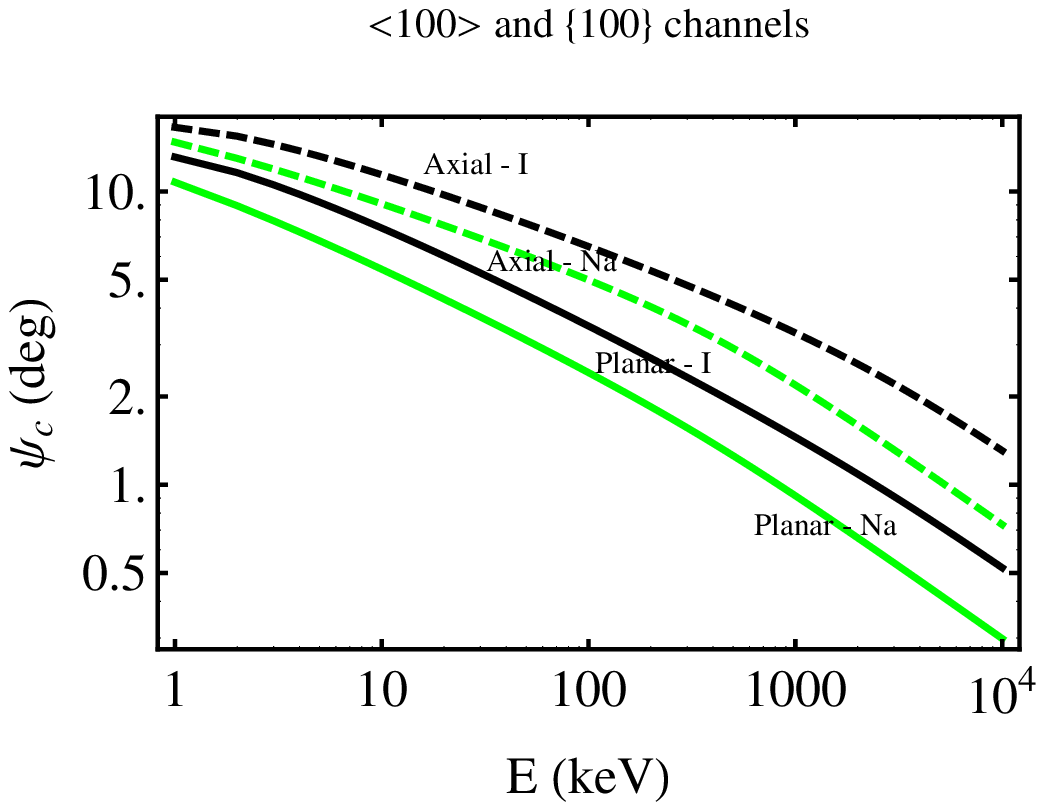,height=160pt}\\
        \vspace{-0.5cm}\caption{(a) Static critical distances of approach and  $u_1$ at 20 $^\circ$C and (b) critical channeling angles at 20 $^\circ$C with $c_1=c_2=1$ as a function of the energy of propagating Na (green/gray) and I (black) ions in the $<$100$>$ axial and \{100\} planar channels. Here $d_{\rm ach}=d_{\rm pch}$.}%
	\label{rc-psic-MV-100}}

\FIGURE{\epsfig{file=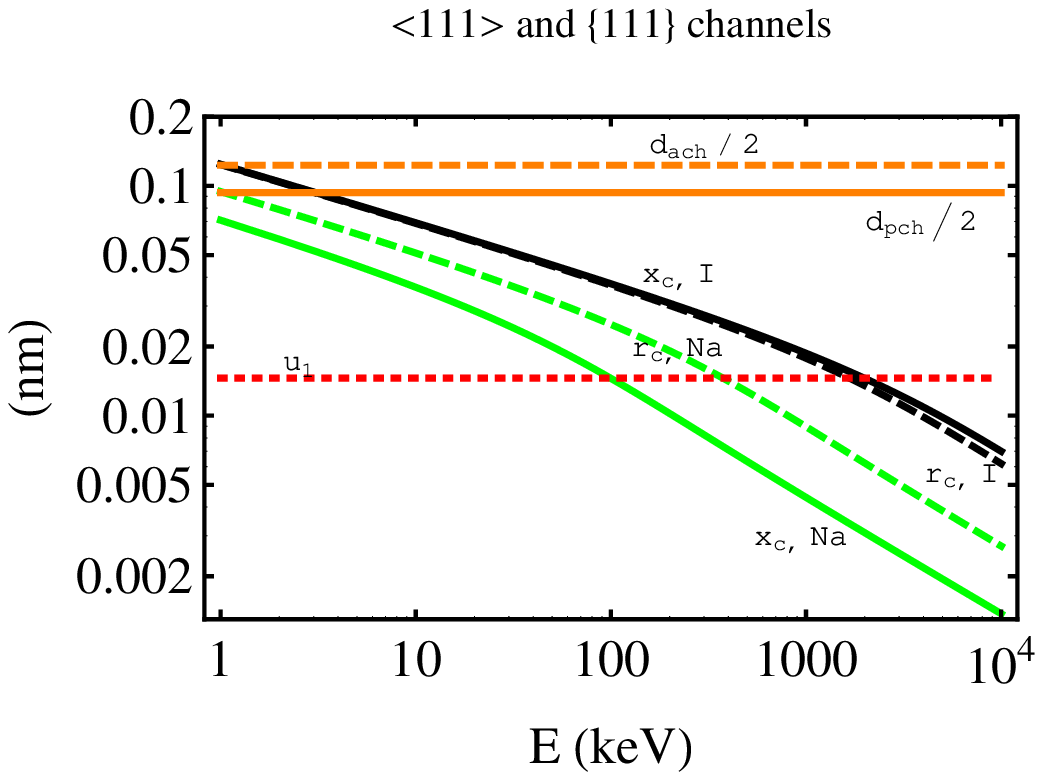,height=160pt}
\epsfig{file=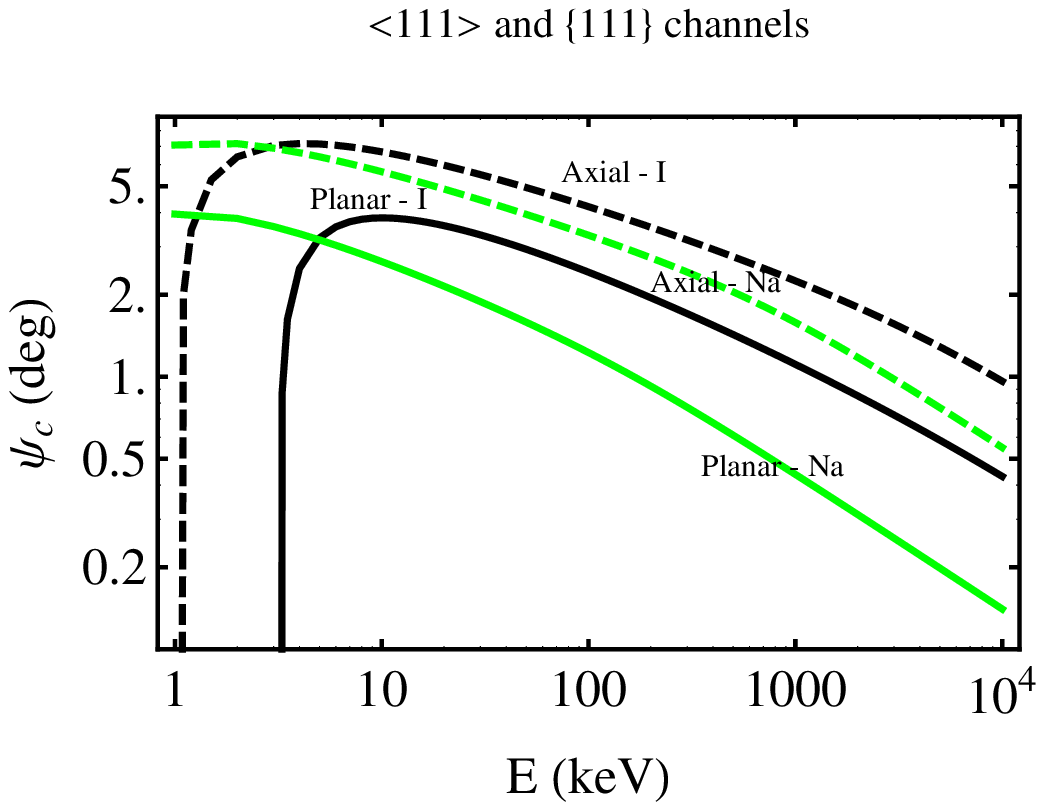,height=160pt}\\
        \vspace{-0.5cm}\caption{Same as Fig.~\ref{rc-psic-MV-100} but for the $<$111$>$ axial and \{111\} planar channels. Here $d_{\rm ach}\not= d_{\rm pch}$.}%
	\label{rc-psic-MV-111}}
\FIGURE{\epsfig{file=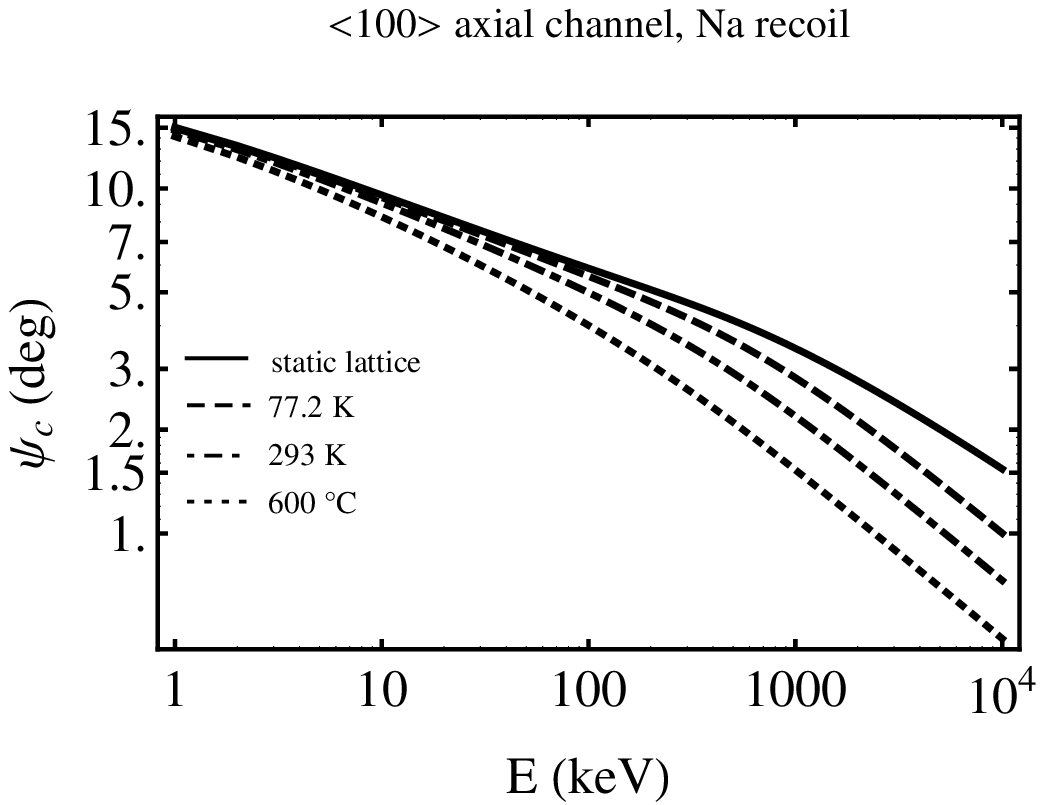,height=170pt}
\epsfig{file=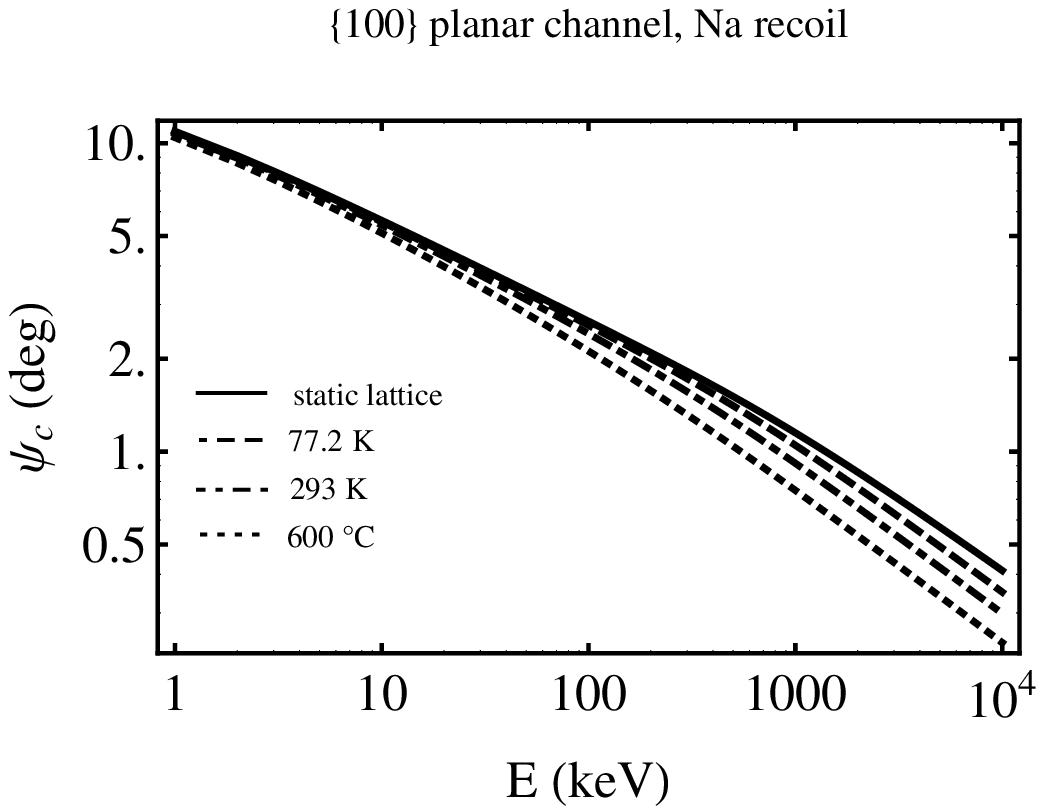,height=170pt}\\
        \vspace{-0.5cm}\caption{Static and T corrected critical angles as a function of energy of Na recoil for T=77.2 K, T=20 $^{\circ}$C, and T=600 $^{\circ}$C for (a) $<$100$>$ axial and (b) \{100\} planar channels.}%
	\label{T-depent-psic}}

In principle there are modifications to the continuum potentials due to thermal effects, but we are going to take into account thermal effects  in the crystal through a modification of the critical distances which was  found originally by Morgan and Van Vliet~\cite{Morgan-VanVliet}  and later by Hobler~\cite{Hobler} to provide good agreement with simulations and data. For axial channels it consists of taking the temperature corrected critical distance $r_c(T)$  to be,
\begin{equation}
r_c(T)= \sqrt{r^2_c(E) + [c_1 u_1(T)]^2},
\label{rcofT}
\end{equation}
where the dimensionless factor $c_1$ in different references is a number between 1 and 2 (see  e.g. Eq.~2.32 of Ref.~\cite{VanVliet} and
Eq.~4.13 of the 1971 Ref~.\cite{Morgan-VanVliet}).

For planar channels the situation is more complicated, because some references give a linear and other a quadratic relation between $x_c(T)$ and $u_1$. Following Hobler~\cite{Hobler} we use an equation similar to that for axial channels,
\begin{equation}
x_c(T)= \sqrt{x^2_c(E) + [c_2 u_1(T)]^2},
\label{xcofT}
\end{equation}
where again $c_2$ is a number between 1 and 2 (for example Barret~\cite{Barrett:1971} finds $c_2 = 1.6$ at high energies, and Hobler~\cite{Hobler}  uses $c_2 = 2$). We will mostly use $c_1=c_2=1$ in the following, to try to produce upper bounds on the channeling fractions.

Using the $T$-corrected critical distances $r_c(T)$ and $x_c(T)$ instead of the static lattice critical distances $r_c$ and $x_c$ in Eqs.~\ref{rcrit} and \ref{ourxcrit}, we obtain the $T$-corrected critical axial and planar angles.

The static axial and planar critical  distances  are presented in Figs.~\ref{rc-psic-MV-100}(a) and  \ref{rc-psic-MV-111}(a) for the 100 and 111 channels, respectively, together with the amplitude of thermal vibrations $u_1$ at 20 $^\circ$C. Figs.~\ref{rc-psic-MV-100}(b) and  \ref{rc-psic-MV-111}(b)  show the  temperature corrected axial and planar  critical angles at 20 $^\circ$C  (with $c_1=c_2=1$) for the same channels as functions of energy of the traveling Na and I ions. We can clearly see in Fig.~\ref{rc-psic-MV-111} that the critical angles become zero at low enough energies (for which the critical distance of approach should be larger than the radius of the channel) indicating the range of energies for which no channeling is possible. Figs.~\ref{T-depent-psic}(a) and \ref{T-depent-psic}(b) show the static and $T$-corrected critical angles at several temperatures for traveling Na ions in the 100 axial and planar channels respectively.

 \section{Channeling of incident particles}

 The channeling of ions in a crystal depends not only on the angle their initial trajectory makes with  strings or planes in the crystal, but also on their initial position. Ions which start their motion close to the center of a channel,  far from a string or plane, where they make an angle $\psi$ or $\psi^p$ respectively, defined in Eqs.~\ref{eq:consetrans} and \ref{planarEperp}, are channeled if the angle  is smaller than a critical angle (as explain earlier) and are not  channeled otherwise. Particles which start their motion in the middle of a channel (as opposed to  a lattice site) must be incident upon the crystal (thus the title of this section).

Here we show that  to a good approximation we can use analytic calculations and reproduce the channeling fraction in NaI presented in Ref.~\cite{Bernabei:2007hw}. It must be noticed, however, that in Ref.~\cite{Bernabei:2007hw} the channeling fraction is computed as if the Na or I ions  started their motion  already within a channel (in fact close to the middle of the channel, where they assume the potential to be negligible),  instead of starting from crystal lattice sites, as is the case in direct dark matter detection. Thus these calculations do not apply to direct dark matter detection experiments.

 Ref.~\cite{Bernabei:2007hw}  considers only a static lattice (i.e. no temperature effects taken into account)  and  the condition for axial channeling they use is $\psi < \psi_2$~\cite{Lindhard:1965}, where $\psi_2$ is defined in Eq.~\ref{psi2}. The equivalent condition for planar channeling used in Ref.~\cite{Bernabei:2007hw} is
\begin{equation}
 \psi^p <  \theta_{pl}=a\sqrt{Nd_p}\left(Z_1 Z_2 e^2/Ea\right)^{1/3}.
  \end{equation}
  The $\theta_{pl}$ angle is very similar to the planar critical angle found by Matyukhin~\cite{Matyukhin} that we call $\psi_c^{pM}$  and is given in Eq.~\ref{psic-M},  $\theta_{pl}= (6 \pi)^{-1/3}\psi_c^{pM}$.

\FIGURE[h]{\epsfig{file=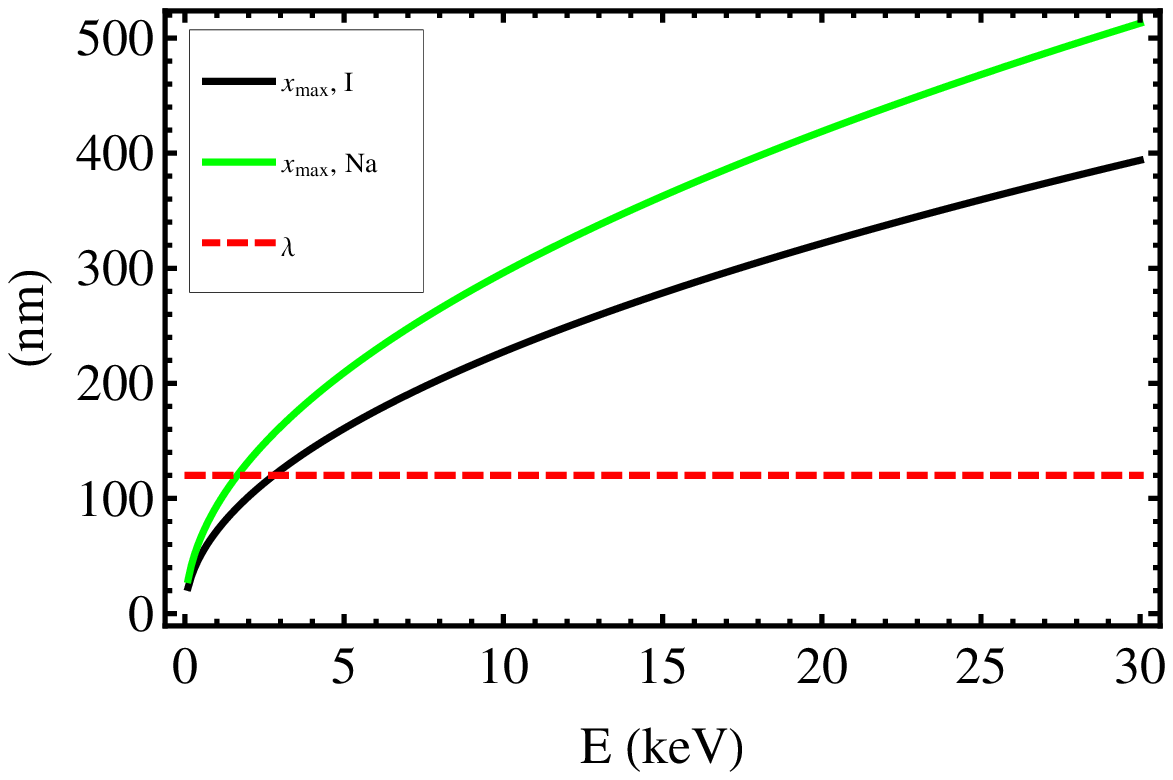,height=210pt}
        \caption{Maximum distance $x_{\rm max}(E)$ traveled  by channeled Na  (green/gray) or I (black) ions in mixed channels of a NaI crystal ($<$100$>$ and $<$111$>$ axial and $\{100\}$ and $\{110\}$ planar channels).}%
	\label{xMax}}

 For an incident angle $\psi$ with respect to each of the channels and an ion energy $E$, the fraction ${\chi}_{\rm inc}(E,\psi)$ of channeled incident ions for axial and planar  channels  is ${\chi}_{\rm inc}=1$ if $\psi$ is smaller than the critical angle for the corresponding channel and zero otherwise.  The NaI structure and all the different channels are explained in Appendix A.

 To find the total fraction $P_{\rm inc}$ of channeled incident nuclei, we average ${\chi}_{\rm inc}$ over  the incident direction $\hat{\textbf{q}}$,
 \begin{equation}
 P_{\rm inc}(E)=\frac{1}{4 \pi}\int{{\chi}_{\rm inc}(E,\hat{\textbf{q}})d\Omega_{q}}.
 \end{equation}
 This integral cannot be solved analytically, so we integrated numerically by performing a Riemann sum once the sphere of directions has been divided using a Hierarchical Equal Area iso-Latitude Pixelization (HEALPix)~\cite{HEALPix:2005} (see Appendix B).

A  channeled ion can be pushed out of a channel by  an interaction with an  impurity such as the atoms of Tl in NaI (Tl). The probability density for an ion  to find  an impurity  after propagating a distance $x$ within the crystal  is
\begin{equation}
p(x)=\frac{1}{\lambda} \exp{\left(-\frac{x}{\lambda}\right)},
\end{equation}
where  $\lambda$  is the average distance between the Tl atoms. We take for $\lambda$ the value used by the DAMA collaboration, i.e. $\lambda=120$ nm~\cite{Bernabei:2007hw}, which according to Ref.~\cite{Drobyshevski:2007zj} corresponds to a molar concentration of 0.0013 Tl atoms for every Na atom.

Here we will simply assume that if a channeled ion interacts with a Tl atom it becomes dechanneled and thus it does not contribute to the fully channeled fraction any longer. We thus neglect  the possibility that after the interaction the ion may reenter into a channel, either the same or another, as we also neglect the possibility that initially non-channeled ions may be  scattered  into a channel. Both effects would increase somewhat the amount of channeled ions, but  we do not have an analytic method of including them in our calculation. Thus the channeled fraction is simply reduced by the probability that the ion does not interact with a Tl atom,
\begin{equation}
P^{ch}(E)=\exp{\left(-\frac{x_{\rm max}(E)}{\lambda}\right)}\, P_{\rm inc}(E) .
\label{Pfully-channeled}
\end{equation}
Here $x_{\rm max}(E)$ is the range of the  propagating ion, i.e.\ the maximum distance a channeled ion with initial energy $E$ can propagate along the channel. Within the channel the ion looses energy into electrons. We use the Lindhard-Scharff~\cite{Lindhard-Scharff, Dearnaley:1973} model of electronic energy loss, valid for energies $E < (M_1/2)Z_1^{4/3} v_0^2$, where $v_0={e^2}/{\hbar} = 2.2 \times 10^8$ cm$/$sec is the Bohr's velocity~\cite{Lindhard:1965}. $M_1$ and $Z_1$ are the mass and charge of the propagating ion. This model is valid for $E< 14.3$ MeV for Na and  $E< $ 646.4 MeV for I in NaI.
\FIGURE[h]{\epsfig{file=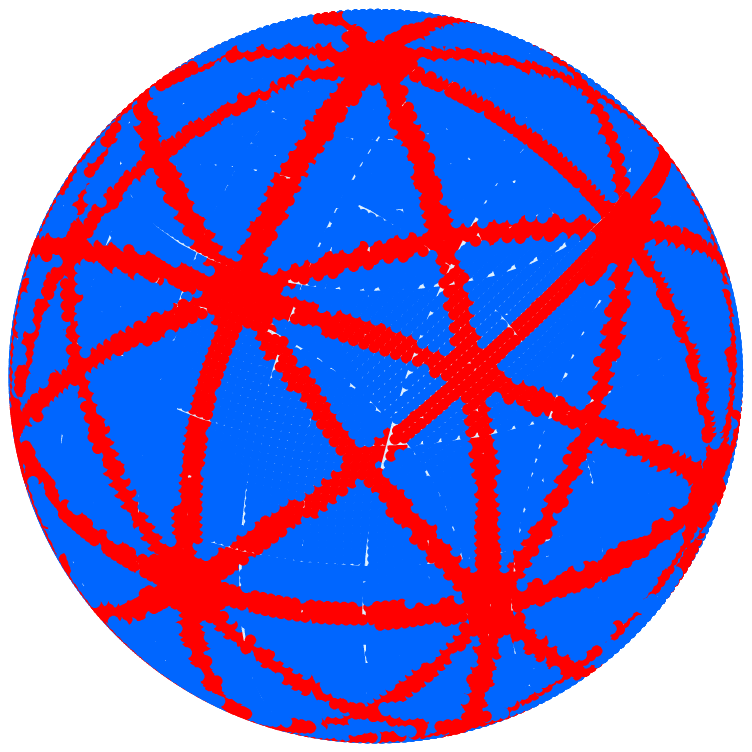,height=190pt}
\epsfig{file=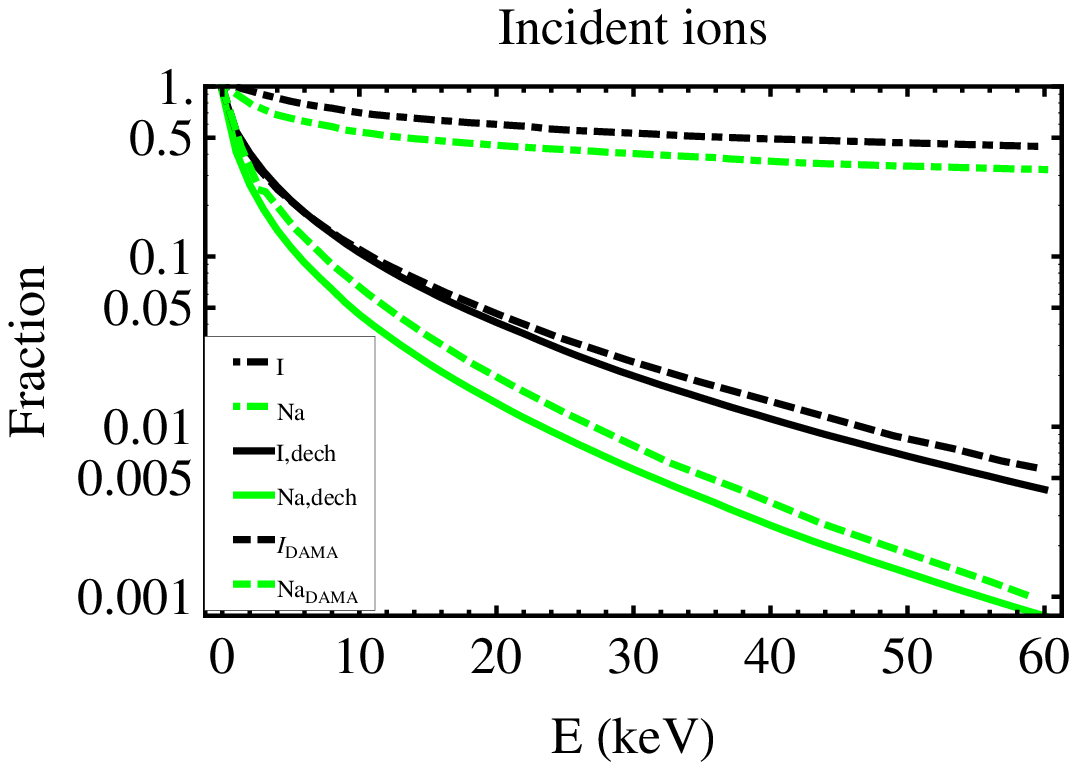,height=160pt}
        \caption{(a) Channeling fraction for a 50 keV Na ion in different directions plotted on a sphere using the HEALPix pixelization:  probability equal to one in red, and probability equal to zero in blue. (b) Fraction of channeled incident  I (black) and Na (green/gray) ions as a function of their incident energy $E$ with the static lattice without  (dot dashed lines) and with (solid lines) dechanneling due to interactions with Tl impurities.  The results of DAMA are also included (dashed lines).}%
	\label{DAMA}}
In this model the energy $E(x)$ as a function of the propagated distance $x$ and the initial energy $E$
 is the  solution of the following energy loss equation~\cite{Dearnaley:1973}
\begin{equation}
-\frac{dE}{dx}=Kv,
\end{equation}
where $v=\sqrt{2E/M_1}$ is the ion velocity and $K$ is the function
\begin{equation}
K=\frac{\xi_e 8 \pi e^2 N a_0 Z_1 Z_2}{ \left(Z_1^\frac{2}{3}+Z_2^\frac{2}{3}\right)^{\frac{3}{2}}v_0}.
\end{equation}
Here $\xi_e$ is a dimensionless constant of the order of $Z_1^{\frac{1}{6}}$~\cite{Dearnaley:1973}, $N$ is the number of atomic centers per unit volume and $a_0\simeq 0.53$ {\AA} is the Bohr radius of the hydrogen atom.
Explicitly, an ion with initial energy $E$ at $x=0$ has energy
\begin{equation}
E(x)=E\left(1-\frac{x}{x_{\rm max}}\right)^2
\label{Ex}
\end{equation}
 after traveling a distance $x$. The range of the propagating ion is
\begin{equation}
x_{\rm max}(E)=\frac{\sqrt{2M_1E_R}}{K}.
\label{xmax}
\end{equation}
 Fig.~\ref{xMax} shows the maximum distance $x_{\rm max}$ traveled  by channeled Na or I ions in mixed channels of an NaI crystal ($<$100$>$ and $<$111$>$ axial and $\{100\}$ and $\{110\}$ planar channels). The average distance $\lambda$ between Tl atoms is also shown.

Fig.~\ref{DAMA}(a) shows the axial and planar channels of the NaI crystal in the HEALPix pixelization of the sphere for  incoming  Na ions with an energy of 50 keV:
 red points indicate a channeling probability of 1 (when the incident angle is smaller than the critical angle with respect to any axial or planar channel) and blue points indicate a channeling probability of zero (when the incident angle is larger than the critical angle). We include here only the channels with lower crystallographic indices, i.e. 100, 110 and 111, which provide the dominant contribution to the channeling fraction,  as is also done in Ref.~\cite{Bernabei:2007hw}.  Fig.~\ref{DAMA}(b) shows  the
 fraction of  incident  I (black lines) or Na (green/gray lines) ions as a function of their incident energy $E$ using the static lattice (solid lines). For comparison,
 Fig.~\ref{DAMA}(b) also shows the channeling fraction obtained by DAMA (dashed lines). Good agreement with the  channeling fractions of DAMA is achieved only when dechanneling due to the interaction with Tl impurities is included.

\section{Channeling of recoiling lattice nuclei}

The recoiling nuclei start initially from lattice sites (or very close to them), thus blocking effects  are important. In fact, as argued originally by Lindhard~\cite{Lindhard:1965}, in a perfect lattice and in the absence of energy-loss processes the probability that a particle starting from a lattice site is channeled would be zero. The argument uses statistical mechanics in which the probability of particle paths related by time-reversal is the same. For example, in optics  if a source of radiation and a point of observation are interchanged, the intensity of the light measured at the new place of observation is the same as the old.  Thus the probability of an incoming ion to have a particular path within the crystal is the same as the probability of the same ion to move backwards along the same path~\cite{Gemmell:1974ub}. This is what Lindhard called the ``Rule of Reversibility."

Using this rule, since the probability of an incoming channeled ion to get very close to a lattice site is zero (otherwise it would suffer a large angle scattering and it would not be channeled), the probability of the same ion to move in the time-reversed path,  starting at a nuclear site and ending inside a channel, is zero too. However, any departure of  the actual lattice from a perfect lattice, for example due to vibrations of the atoms in the lattice, would violate the conditions of this argument and allow for some of the recoiling lattice nuclei to be channeled.

The channeling of particles emitted at lattice sites  due to lattice vibrations, such as protons scattered at large angles, was measured and already understood in the 70's. Komaki {\it et al.}~\cite{Komaki-et-al-1971} in a 1971 paper titled
 ``Channeling Effects in the Blocking Phenomena"  observed channeling  of protons scattered at large angles within  thin Si and Ge  crystals and explained it as due to the fact that the scattering or emitting lattice atom is not exactly at the lattice site because of thermal vibrations." They fit their data using the model presented by Komaki and Fujimoto~\cite{Komaki:1970} one year earlier.

We now estimate the channeling fractions using the formalism presented so far.

\subsection{Channeling fraction for each channel}

We need to know how probable it is for the recoiling nucleus  to be at a particular distance $r$ from its equilibrium position in a crystal row when it collides with a WIMP.  The probability distribution function $g(r)$ of the perpendicular distance to the row of the colliding atom due to thermal vibrations can be represented by a two-dimensional Gaussian (as done by Lindhard and many others~\cite{Gemmell:1974ub}, the relevant vibrations being in the plane orthogonal to the row),
\begin{equation}
g(r)=\frac{r}{u_1^2}\exp{(-r^2/2u_1^2)}.
\label{gofr}
\end{equation}
The one dimensional vibration amplitude $u_1$ is given in Eq.~\ref{vibu1}.

The channeled fraction $\chi_{\rm axial}(E, \hat{\textbf{q}})$ of nuclei with recoil energy $E$ moving initially in the direction $\hat{\textbf{q}}$ making an  angle $\phi$ with respect to the axis is given by the fraction of nuclei which can be found at a distance $r$ larger than a minimum distance $r_{i,\rm min}$ from the row at the moment of collision, determined by the critical distance of approach as shown in the next subsection,
\begin{equation}
\chi_{\rm axial}(E, \phi)=\int_{r_{i,\rm min}}^{\infty}{dr g(r)}=\exp{(-r_{i,\rm min}^2/2u_1^2)}.
\label{chiaxial}
\end{equation}
Note that here we are approximating the upper limit of the integral of $g(r)$ with  $\infty$, instead of  the radius of the axial channel $d_{\rm ach}/2$. This is a good approximation because $d_{\rm ach}/2 \simeq 10 a$ or more and the integral is dominated by the values of $g(r)$ close to $u_1 \ll 2a$.

If $\phi>\psi_c$ no channeling can occur and $\chi_{\rm axial}(E, \phi)=0$. This can easily be seen from Eqs.~\ref{E-perp-HP},  \ref{r*definition},  \ref{eq:consetrans}  and  \ref{defrcrit}, taking into account that $U(r_i + d \tan\phi) \geq U(r_c)$.

For a planar channel, the Gaussian thermal distribution for the planar potential is one-dimensional (the relevant vibrations occurring perpendicularly to the plane),
\begin{equation}
g(x)=  (2 \pi u_1^2)^{-1/2} \exp(-x^2/2u_1^2) .
\label{gofx}
\end{equation}
This is normalized  to 1 for $-\infty<x<+\infty$. In our calculations we only consider positive values of $x$ for each plane, thus we multiply $g(x)$ by a factor of $2$ to find the fraction of channeled nuclei for a planar channel,
\begin{equation}
\chi_{\rm planar}(E, \phi)=\int_{x_{i,\rm min}}^{\infty}{2g(x) dx}
=\frac{2}{\sqrt{\pi}}\int _{x_{i,\rm min}}^{\infty}{\frac{e^{(-x^2/2u_1^2)}}{\sqrt{2}u_1}dx}\\
=\mathop{\rm erfc}\left(\frac{x_{i,\rm min}}{\sqrt{2}u_1}\right).
\label{chiplanar}
\end{equation}
Here $\phi$ is the angle $\hat{\textbf{q}}$ makes with the plane, defined as the complementary angle to the angle between $\hat{\textbf{q}}$ and the normal to the plane, or as the smallest angle between $\hat{\textbf{q}}$ and vectors lying on the plane. Similar to the axial case, we approximate the upper limit of the integral of $g(x)$ with  $\infty$, instead of  the radius of the planar channel $d_{\rm pch}/2$. This is a good approximation because $d_{\rm pch}/2 \simeq 10 u_1$ or more, and ${\rm erfc}[d_{\rm pch}/(2 \sqrt{2} u_1)]$ is negligible. Also in this case, $\chi_{\rm planar}(E, \phi)=0$ if $\phi$ is larger than the critical channeling angle of the particular channel, i.e. if $\phi>\psi_c^p$.

We conclude this subsection by noticing an important point. In Eq.~\ref{chiaxial}, $r_{i,\rm min}$, which is a function of $r_{c}(T)$, enters exponentially. Thus any uncertainty in our modeling of $r_{c}(T)$ becomes exponentially enhanced in the channeling fraction. The same happens with the dependence of the channeling fraction in Eq.~\ref{chiplanar} on $x_{i,\rm min}$, which depends on $x_c(T)$. This is the major difficulty of the analytical approach we are following.

 \subsection{Minimum initial distance of the recoiling lattice nucleus}

For axial channels, using Eqs.~\ref{E-perp-HP},  \ref{r*definition},  \ref{eq:consetrans}  and  \ref{defrcrit}, we can write the condition for channeling as
\begin{equation}
E \sin^2 \phi + U(r_i + d \tan\phi)= U(r_{\rm min}) < U(r_c(E)).
\label{ChanCond-CA}
\end{equation}
Therefore, the minimum initial distance $r_{i,\rm min}$ is the solution of the equation
\begin{equation}
U(r_{i,\rm min}+ d \tan\phi) = U(r_c(E))-E \sin^2\phi.
\end{equation}
Inverting the function $U(r)$ we obtain
\begin{equation}
r_{i,\rm min}(E,\phi)+ d \tan\phi = U^{-1}[ U(r_c(E))-E \sin^2\phi ].
\end{equation}
The inverse of Lindhard's potential function $U(r)$ is
\begin{equation}
U^{-1}(r) = \frac{C a}{\sqrt{e^{2r/E_1}-1}},
\end{equation}
which together with the expression for $r_c(E)$ in Eq.~\ref{rcrit} yields a fully analytic expression for $r_{i,\rm min}(E,\phi)$,
\begin{equation}
r_{i,\rm min} (E, \phi) = \frac{C a}{\sqrt{\left( 1+\frac{C^2a^2}{r_{c}^2} \right) \, \exp\!\left(-{2 \sin^2\phi}/{\psi_1^2} \right) -1 }} - d \tan\phi.
\end{equation}

\FIGURE{\epsfig{file=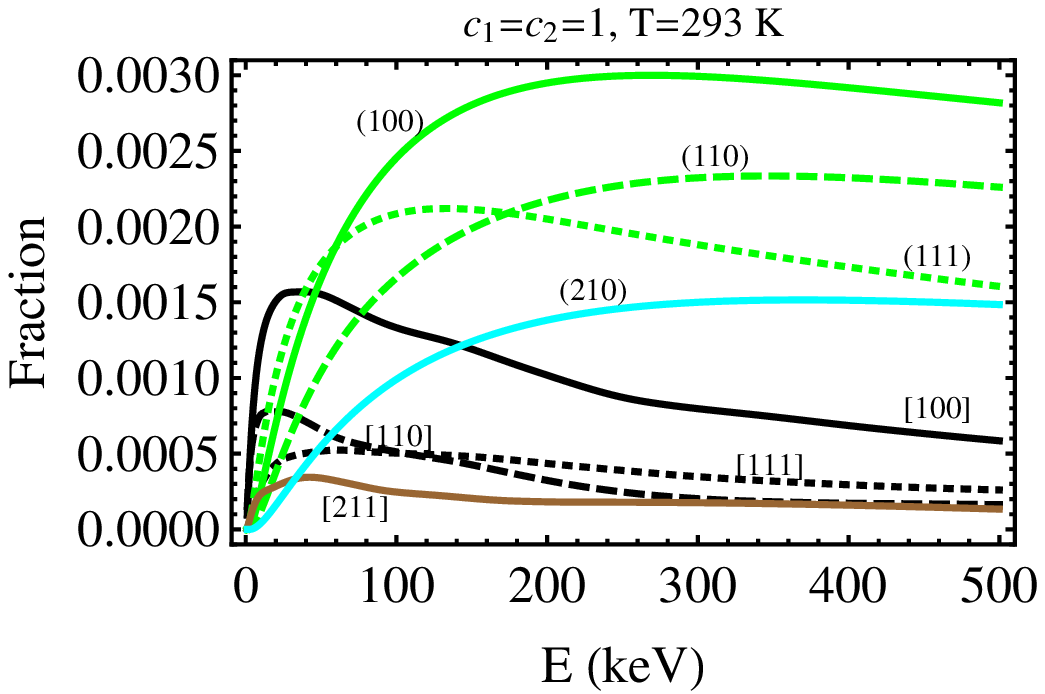,height=210pt}
        \caption{Upper bounds to the channeling fractions of Na recoils for single planar (green/gray lines) and axial (black lines) channels, as function of the recoil energy $E$, for T$=293$ K and $c_1=c_2=1$,  without including dechanneling. Two additional channels not included in the total channeling fractions are also shown: axial $[211]$ (brown line) and planar $(210)$ (cyan line).}%
	\label{OneChannel}}
\FIGURE{\epsfig{file=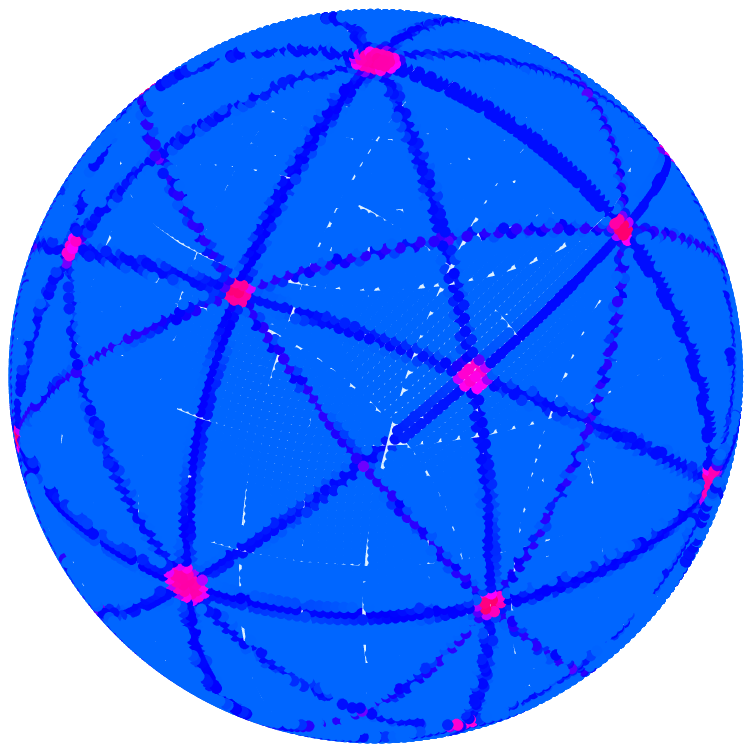,height=170pt}
\epsfig{file=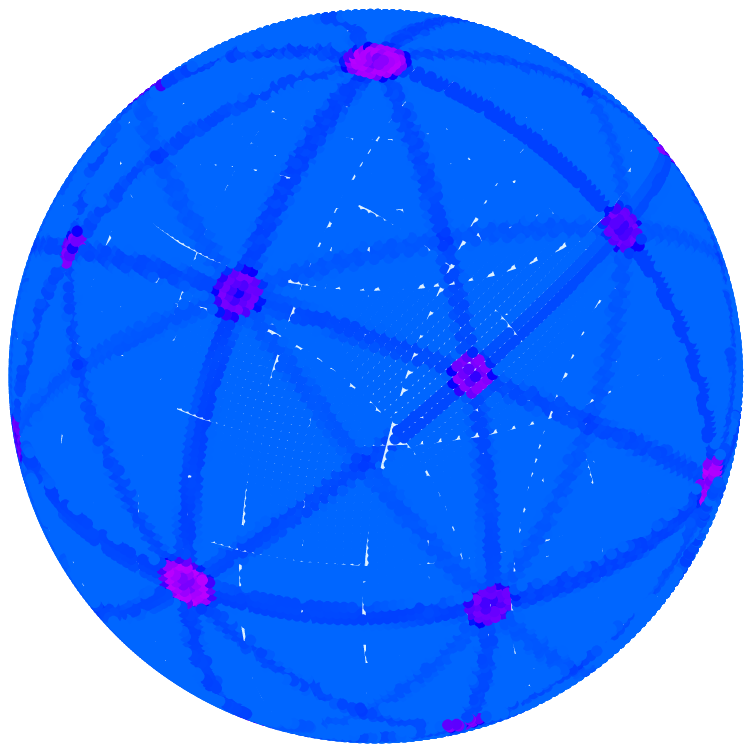,height=170pt}
        \vspace{-0.5cm}\caption{Channeling probability $\chi_{\rm rec}(E, {\bf{\hat{q}}})$ (Eq.~\ref{chirec}) for a 200 keV recoil of (a) an Na ion and (b) an I ion at 20 $^\circ$C with $c_1=c_2=1$ and neglecting dechannneling. The probability is computed for each direction and plotted on a sphere using the HEALPix pixelization. The red, pink, dark blue and light blue colors indicate a channeling probability of 1, 0.625, 0.25 and zero, respectively.}%
	\label{Our-HEALPIX}}

Applying the same arguments  to planar channels, we have
\begin{equation}
U(x_{i,\rm min} + d_p \tan\phi) = U(x_c(E))-E \sin^2\phi .
\end{equation}
For Lindhard's potential, the minimum initial distance is given by
\begin{equation}
x_{i, \rm min}(E, \phi)=\frac{a}{2}\,\frac{C^2-\left[\sqrt{\frac{x_c^2}{a^2}+C^2}-\frac{x_c}{a}-{\sin^2\phi}/{\psi_a^2}\right]^2}{\left[\sqrt{\frac{x_c^2}{a^2}+C^2}-\frac{x_c}{a}-{\sin^2\phi}/{\psi_a^2}\right]} - d_p \tan\phi.
\end{equation}
Here, $x_c(E)$ is found in Eq.~\ref{ourxcrit}.

Fig.~\ref{OneChannel} shows upper bounds to the channeling fractions of Na recoils for individual channels with $c_1=c_2=1$ and T$=293$ K, without including dechanneling. The black and green (or gray) lines correspond to single axial and planar channels respectively. Two additional channels not included in the total channeling fractions are also shown for comparison: the axial $[211]$ (brown line) and the planar $(210)$ (cyan line) channels. The upper bounds of channeling fractions for planar channels are more generous than those of axial channels because of our choice of $x_c$ in Eq.~\ref{ourxcrit}. This does not mean that planar channels are dominant in the actual channeling fractions.

\subsection{Channeling fraction}

The geometric channeling fraction is the fraction of recoiling ions that propagate in the 1st, or 2nd, or \ldots or 26th channel. Here ``geometric'' refers to assuming that the distribution of recoil directions is isotropic. In reality, in a dark matter direct detection experiment, the distribution of recoil directions is expected to be peaked in the direction of the average WIMP flow. For comparison with previous work of others, here we examine this geometrical channeling fraction, and postpone the case of a WIMP wind to another paper \cite{NGG-WIMPwind}.

We include only the channels with lowest crystallographic indices, i.e. 100, 110 and 111, which are in total 26 axial and planar channels, as explained in Appendix A. We have also checked other axial and planar channels, such as the $[211]$ and $(210)$ channels shown in Fig.~\ref{OneChannel}, and found that their contribution to the channeling fractions is negligible (additional planar channels are always less important than the planar channels we keep and the same happens for axial channels).

The probability $\chi_{\rm rec}(E,\hat{\bf q})$ that an ion with initial energy $E$ is channeled in a given direction $\hat{\bf q}$ is the probability that the recoiling ion enters any of the available channels. We compute it using a recursion of the addition rule in probability theory over all axial and planar channels:
\begin{eqnarray}
P(A_1~\text{or}~A_2) &=& P(A_1) + P(A_2) - P(A_1)P(A_2).\nonumber\\
P(A_1~\text{or}~A_2~\text{or}~A_3) &=& P(A_1~\text{or}~A_2) + P(A_3) - P(A_1~\text{or}~A_2)P(A_3).
\label{Prob}
\end{eqnarray}
We continue this recursive computation until we find the probability with which the recoiling ion goes into any of the 26 channels
\begin{equation}
\chi_{\rm rec}(E,\hat{\bf q}) = P(A_1~\text{or}A_2~\text{or \ldots or}~A_{26}).
\label{chirec}
\end{equation}
For each channel $A_k$ ($k=1,\ldots,26$), $P(A_k)= \chi_{{\rm axial}-k}(E, \phi_k)$ or  $P(A_k)= \chi_{{\rm planar}-k}(E, \phi_k)$ for an axial or planar channel, respectively.  Notice that $P(A_k) \neq 0$ only for the channels for which $\phi_k < (\psi_c)_k$, i.e.\ for which the angle that $\hat{\bf q}$ makes with the axis or plane of the channel, respectively, is smaller than the critical angle for the channel.

 Here we are treating channeling along different channels as independent events, so that  the conditional probabilities coincide with the non-conditional probabilities, e.g. $P(A_1|A_2)=P(A_1)$. This is correct for different axial channels, which never overlap, and a good approximation for different planar channels. However, axial channels happen at the crossing of two or more planar channels, thus channeling into axial and planar channels may not be entirely independent. We prove in Appendix D that considering them as independent is, however, a good approximation.

\FIGURE{\epsfig{file=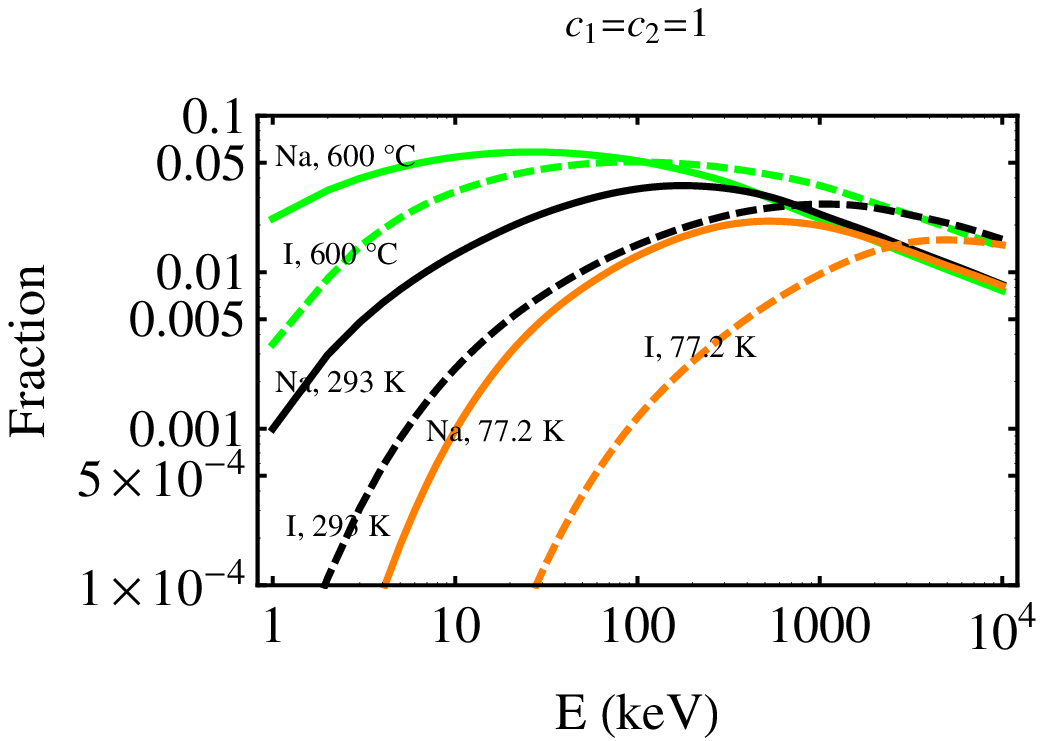,height=160pt}
\epsfig{file=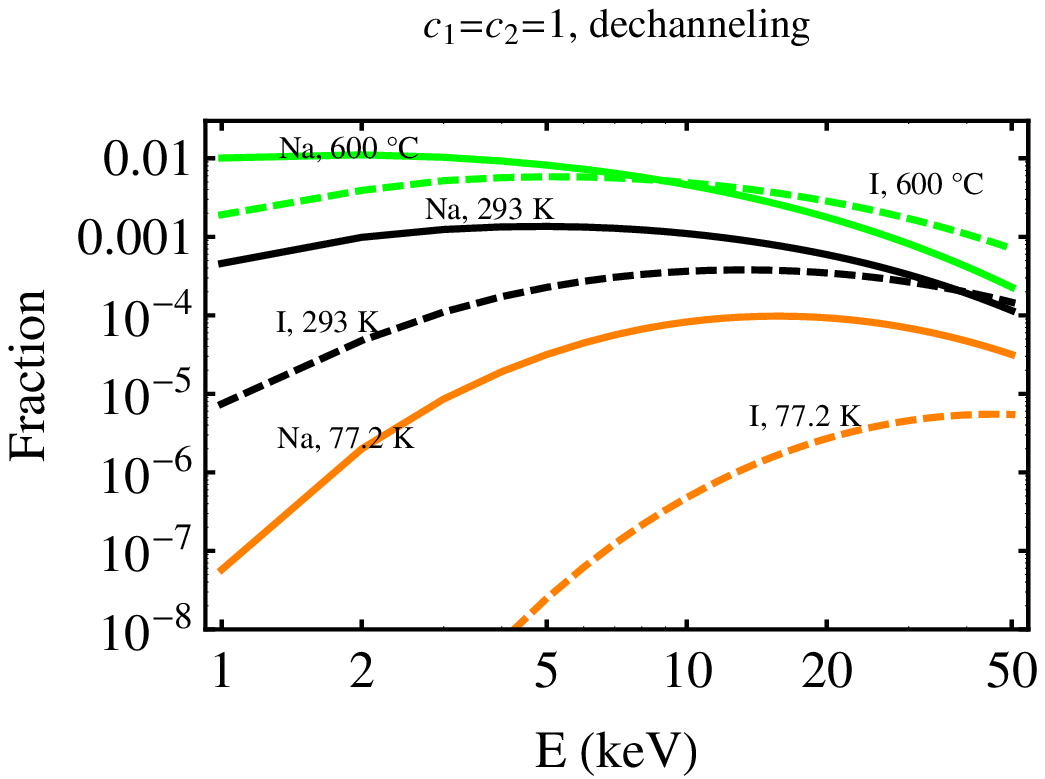,height=160pt}
       \vspace{-0.5cm} \caption{Upper bounds to the channeling fraction of Na and I recoils as a function of the recoil energy $E$ for T=600 $^\circ$C (green/light gray), 293 K (black), and 77.2 K (orange/dark gray)  in the approximation of $c_1=c_2=1$, (a) without and (b) with dechanneling as in Eq.~\ref{Pfully-channeled}.}%
	\label{FracNaI-DiffT-c1}}

\FIGURE{\epsfig{file=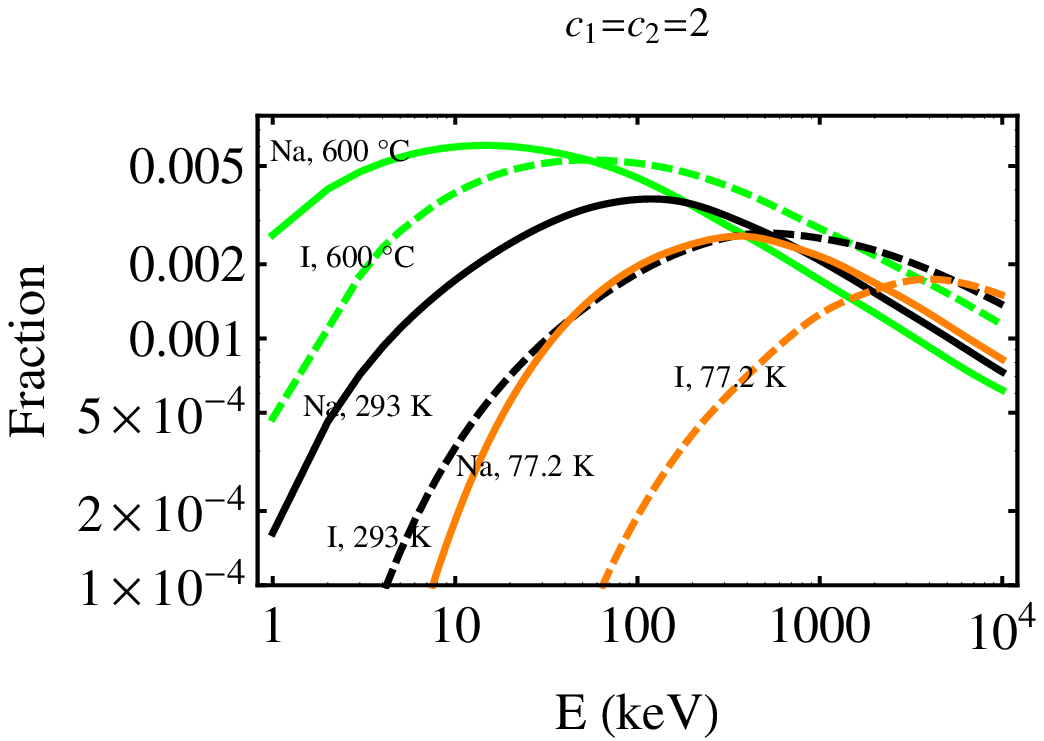,height=160pt}
\epsfig{file=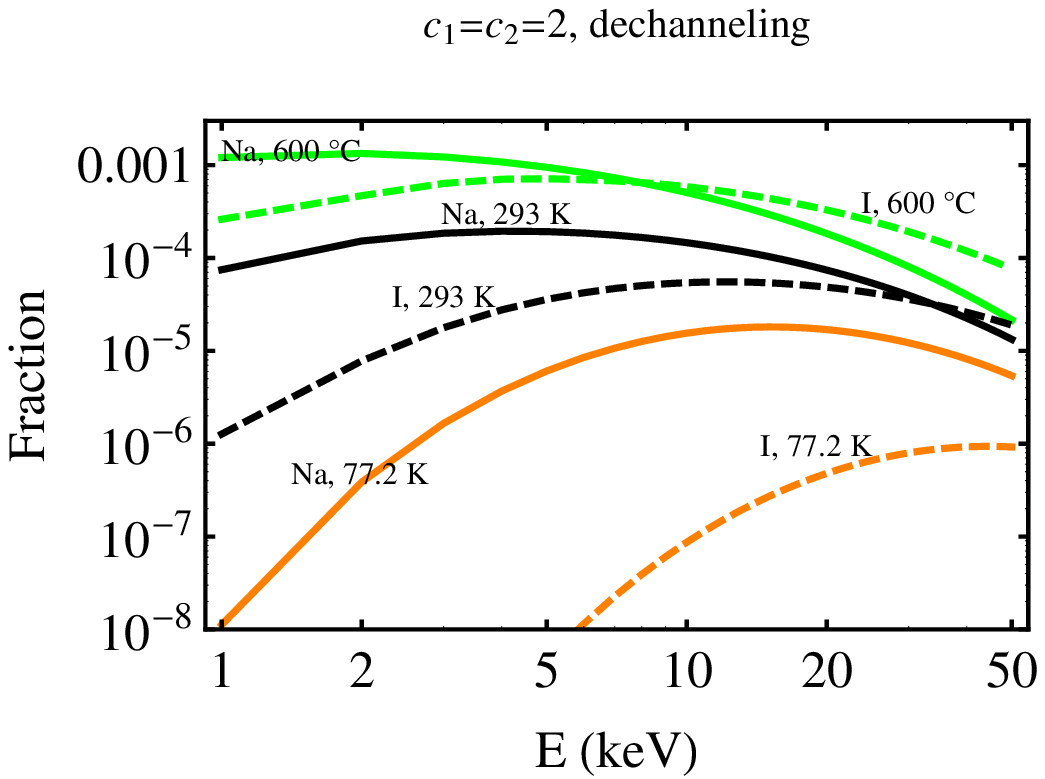,height=160pt}
        \vspace{-0.5cm}\caption{Same as Fig.~\ref{FracNaI-DiffT-c1} but for $c_1=c_2=2$.}%
	\label{FracNaI-DiffT-c2}}

\FIGURE{\epsfig{file=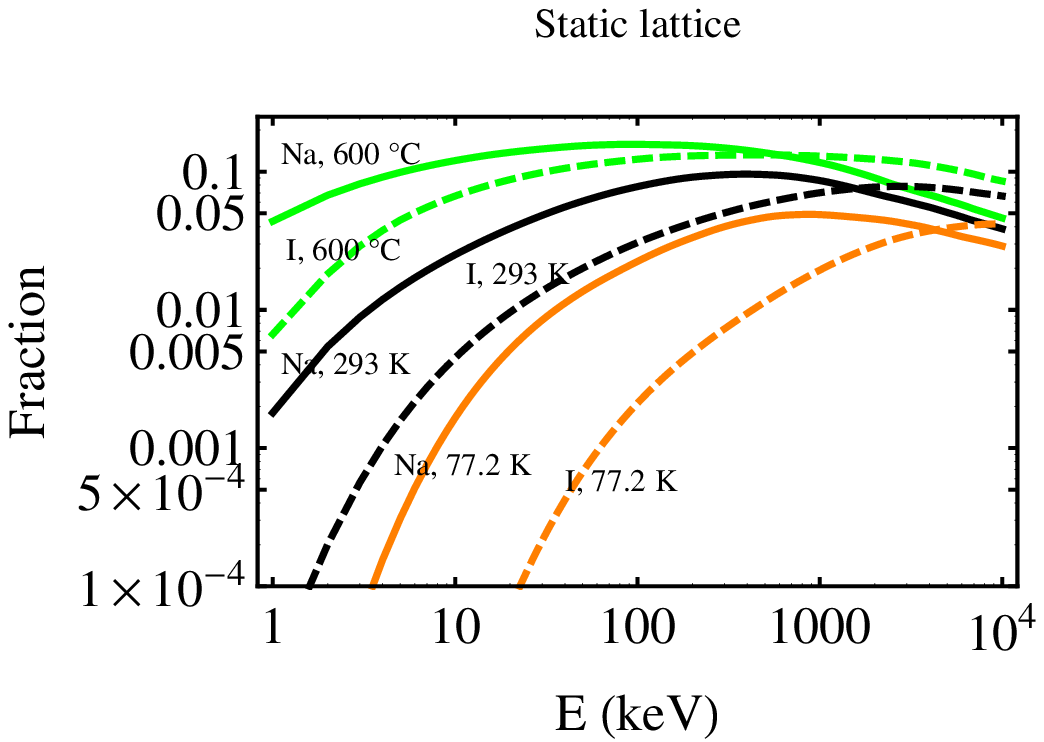,height=160pt}
\epsfig{file=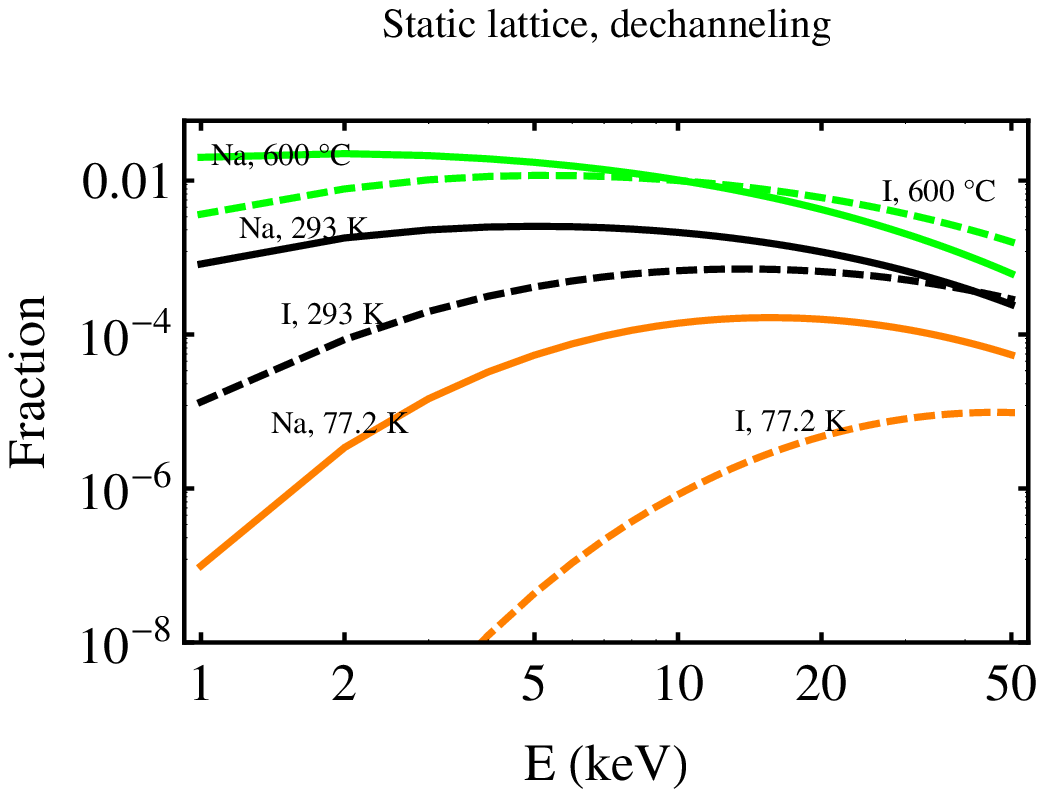,height=160pt}
        \caption{Same as Fig.~\ref{FracNaI-DiffT-c1} but for $c_1=c_2=0$ (static lattice),  provided as an upper bound with respect to any non-zero values of $c_1$ and $c_2$.}%
	\label{FracNaI-DiffT-rigid}}

Fig.~\ref{Our-HEALPIX} shows the channeling probability for an $E= 200$ keV recoil of (a) Na or (b) I at 20 $^\circ$C with $c_1=c_2=1$ and neglecting dechanneling,  computed for each direction $\hat{\bf q}$ and plotted on a sphere using the HEALPix pixelization. The red, pink, dark blue and light blue colors indicate a channeling probability of 1, 0.625, 0.25 and zero, respectively.

To obtain the geometrical channeling fraction, we average the channeling probability $\chi_{\rm rec}(E,\hat{\bf q})$ over the directions $\hat{\bf q}$, assuming an isotropic distribution of the initial recoiling directions $\hat{\bf q}$,
\begin{equation}
P_{\rm rec}(E)=\frac{1}{4\pi}\int{\chi_{\rm rec}(E, \hat{\bf q})d\Omega_q}.
\label{probrec}
\end{equation}
This integral is computed using HEALPix~\cite{HEALPix:2005} (see Appendix B).

The channeling fraction as a function of recoil energy is shown in Figs.~\ref{FracNaI-DiffT-c1} and \ref{FracNaI-DiffT-c2} with $c_1=c_2=1$ and $c_1=c_2=2$, respectively. These curves include thermal effects in the lattice at various crystal temperatures, and are shown (a) without  and (b) with dechanneling according to Eq.~\ref{Pfully-channeled}.  Note that  the dechanneling as included here is possibly too extreme, since it does not allow for the possibility of an ion reentering a channel (the original or another one) after a collision with a Tl impurity. As shown in Fig.~\ref{FracNaI-DiffT-c1}(a), the maximum of the channeling probability for Na recoils is at energies close to 200 keV.

Notice that while the effect of increasing temperatures on the initial position of the recoiling nucleus makes the channeling fractions larger (the recoiling nucleus can be initially further out from the string, i.e.\ $u_1$ in Eqs.~\ref{chiaxial} and \ref{chiplanar}  increases), the effect of increasing temperatures on the other lattice atoms is to increase the  critical distances, which makes the channeling fractions smaller ($r_{i, \rm min}$ and $x_{i, \rm min}$ in Eqs.~\ref{chiaxial} and \ref{chiplanar} increase). Figs.~\ref{FracNaI-DiffT-c1} and \ref{FracNaI-DiffT-c2} show that in our calculations the first effect is almost always dominant, except that for $c_1=c_2=2$ at some energies the temperature effects  in the lattice are larger (see the left panel of Fig.~\ref{FracNaI-DiffT-c2}, where some fractions are smaller at higher temperatures). Neglecting the temperature  effects in the lattice, by
setting $c_1=c_2=0$, but including the thermal vibrations of the nucleus that is going to recoil, we obtain the largest estimates for the channeling fractions. Although it is physically inconsistent to take only the temperature effects on the initial position of the recoiling nuclei but not on the lattice, this was done by Lindhard~\cite{Lindhard:1965} and Andersen~\cite{Andersen:1967} early on, and we do it here because it provides the  most generous upper bound on the channeling fraction (any  non-zero value of $c_1$ or $c_2$ would lead to smaller fractions).

\section{Conclusions}

We have studied the channeling of ions recoiling after collisions with WIMPs inside an NaI (Tl) crystal.   Channeled ions move within the crystal along symmetry axes and planes and suffer a series of small-angle scatterings  that maintain them in the open ``channels"  between the rows or planes of lattice atoms and thus penetrate much further into the crystal than in other directions.  In order for the scattering to happen at small enough angles, the propagating ion must not approach a row or plane closer than a critical distance $r_c$ for axial or $x_c$ for planar channels. For a ``static lattice" that here means a perfect lattice in which all vibrations are neglected, $r_c$ and $x_c$ are given in Eqs.~\ref{rcrit} and \ref{ourxcrit}.

The channeling of ions in a crystal depends not only on the angle their initial trajectory makes with  rows or planes in the crystal, but also on their initial position. Ions which start their motion close to the center of a channel,  far from a row or plane, at an initial angle $\psi$ or $\psi^p$ (see Eqs.~\ref{eq:consetrans} and \ref{planarEperp}), are channeled if the initial angle  is smaller than the critical angle in Eqs.~\ref{psicritaxial} and \ref{psicritplanar}, respectively, and are not  channeled otherwise. We have found that the channeling of lattice ions recoiling after a collision with a WIMP is very different from the channeling of incident ions, and that the  fraction of recoiling lattice ions that are channeled is smaller.

The nuclei ejected from their lattice sites by WIMP collisions are initially part of a row or plane. They start from lattice sites or very close to them, thus blocking effects  are important. In fact, as argued originally by Lindhard~\cite{Lindhard:1965}, in a perfect lattice and in the absence of energy-loss processes, the probability that a particle starting from a lattice site is channeled would be zero.  This is what Lindhard called the ``Rule of Reversibility." However, any departure of  the actual lattice from a perfect lattice due to vibrations of the atom,  which are always present, violate the conditions of this argument and allow for some of the recoiling lattice nuclei to be channeled. Thus, the channeling fraction of recoiling ions is very temperature dependent.

Due to vibrations  in the crystal, the atom that interacts with a WIMP may be displaced from its position in a perfect lattice with a probability given in Eqs.~\ref{gofr} and \ref{gofx}. It is this displacement which allows for a non-zero probability of channeling, given in Eqs.~\ref{chiaxial} and \ref{chiplanar}. At high temperatures, the atoms vibrate with larger amplitudes, the recoiling ion can start further away from a row or plane (i.e\ $u_1$ in Eqs.~\ref{chiaxial} and \ref{chiplanar}  is larger), and the channeling fractions increase.
 This is illustrated in Figs.~\ref {FracNaI-DiffT-c1}, \ref{FracNaI-DiffT-c2} and \ref{FracNaI-DiffT-rigid}.  These three figures differ because the lattice vibrations (of all the other atoms in the crystal, besides the recoiling one) increase the critical distances of approach  and reduce the critical angles for channeling (unless $c_1=c_2=0$), which in turn decreases the channeling fractions as the temperature increases ($r_{i, \rm min}$ and $x_{i, \rm min}$ in Eqs.~\ref{chiaxial} and \ref{chiplanar} increase).

 Without better data or simulations of Na and I ions propagating in a NaI crystal, the best we can do  is to write the $T$-corrected minimal distances of approach $r_c(T)$ and $x_c(T)$ as functions of the two parameters $c_1$ and $c_2$. The values of $c_1$ and $c_2$ we found in the literature for different materials and propagating ions are between 1 and 2 (see Eq.~\ref{rcofT} and \ref{xcofT}).

\FIGURE{\epsfig{file=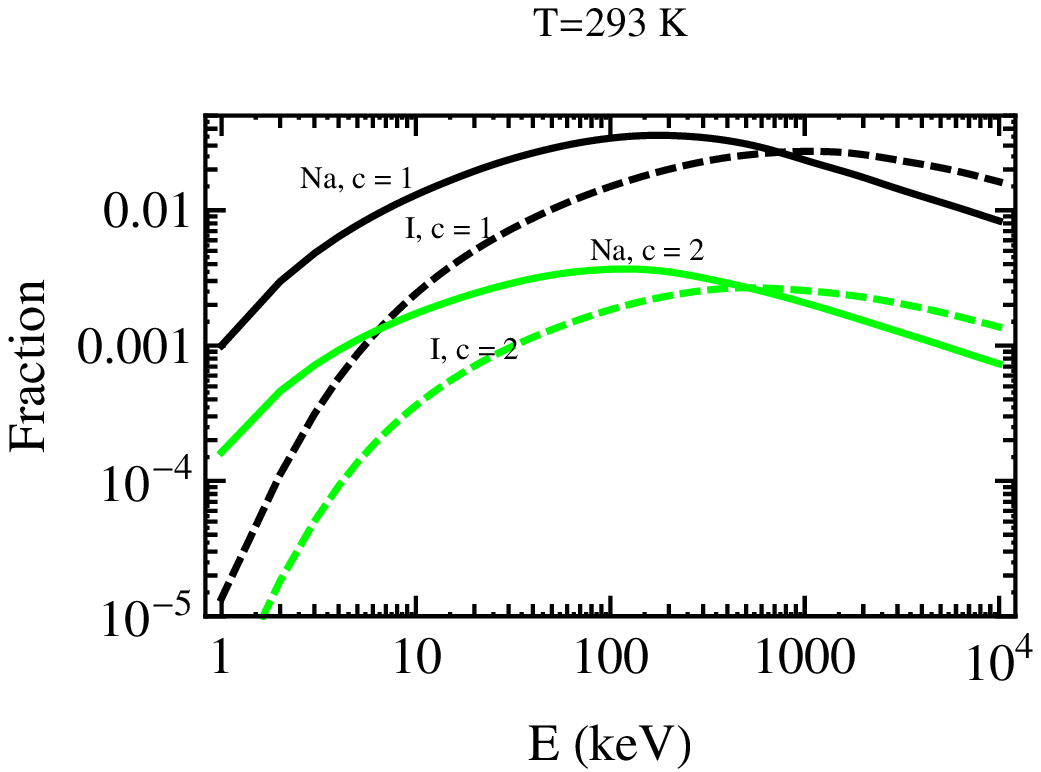,height=160pt}
\epsfig{file=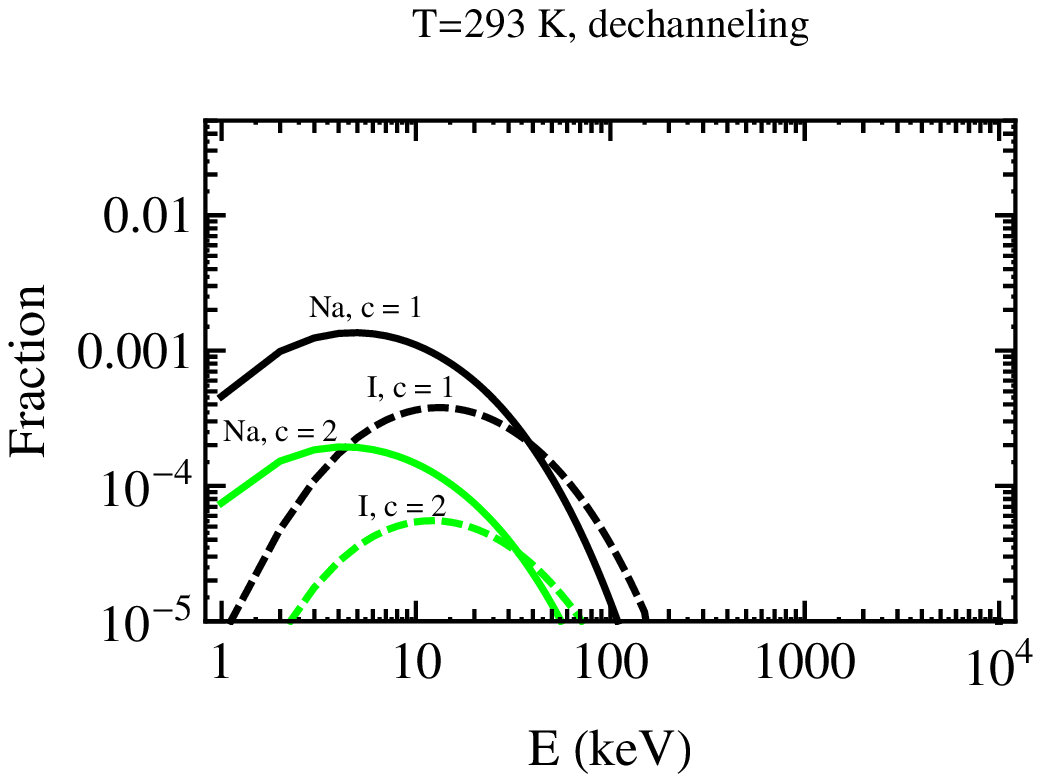,height=160pt}
        \caption{Channeling fractions at T=293 K for Na (solid lines) and I (dashed lines) ions for $c=c_1=c_2=1$ (black) and $c=c_1=c_2=2$ (green/gray) cases (a) without and (b) with dechanneling included as in Eq.~\ref{Pfully-channeled}.}%
	\label{FracNaI-Final}}
The coefficients $c_1$ and $c_2$ determine the dependence of the critical distances of approach on the vibration amplitude $u_1(T)$. This dependence is important at relatively large energies, when
the static critical distances $r_c$ and $x_c$ become small with respect to $u_1$.

We are already providing upper limits to the channeling fraction due to our choice of $x_c$ (see discussion after Eq.~\ref{ourxcrit}). In Fig.~\ref{FracNaI-DiffT-rigid} we decided to show absolute upper bounds to this probability in the unrealistic case in which temperature effects are only taken into account in the vibrations of the atom interacting with the WIMP, but not on the other atoms in the lattice (so $c_1=c_2=0$).  More realistic upper bounds to the channeling fraction (without dechanneling) are given in Fig.~\ref {FracNaI-DiffT-c1}, in which $c_1=c_2=1$. The 20 $^\circ$C curve from this figure is displayed again in Fig.~\ref{FracNaI-Final}, in which we show what we consider to be our best results. The larger are  $c_1$ and $c_2$, the larger are the temperature corrected critical channeling distances, the smaller are the temperature corrected critical channeling angles, and thus the smaller are the resulting channeling fractions.  If $c_1=c_2=2$, found by Hobler to be necessary to reproduce the measured critical angles at $E< 100$ keV in Si (see our companion paper devoted to Si and Ge~\cite{NGG-SiandGe}), the channeling fractions are smaller, as shown in Fig.~\ref{FracNaI-DiffT-c2} from which the 20 $^\circ$C curve is copied in Fig.~\ref{FracNaI-Final}.

 Fig.~\ref{FracNaI-Final}(a) shows what we consider to be our main predictions for the range expected as an upper limit to the channeling fraction in NaI (Tl), if dechanneling is ignored.  Dechanneling happens when the channeled ion encounters impurities or defects. Fig.~\ref{FracNaI-Final}(b) shows the channeling fraction reduced by the probability of the channeling ion to not interact with a Tl atom
 (see Eq.~\ref{Pfully-channeled}). This way of taking into account dechanneling may be too extreme, as it neglects the probability that the ion after the collision with a Tl atom may reenter a channel (either the same channel or another one) and be again channeled.

 As we see in Fig.~\ref{FracNaI-Final}(a), neglecting interactions with Tl atoms, the channeling fraction is never larger than 5\% and the maximum happens at 100's of keV. This maximum occurs because the critical distances decrease with the ion energy $E$, making channeling more probable, and the critical angles also decrease with $E$, making channeling less probable. All this is without dechanneling.

With dechanneling, the probability that the channeled ion does not interact  with a Tl atom decreases with energy (since more energetic ions propagate further within channels). Thus, interactions with Tl atoms decrease the channeling fraction at high energies. The simple extreme model of dechanneling we use in this paper predicts much smaller fractions, at most in  the 0.1\% level, with the maximum shifted to small energies, less than 10 keV (see Fig.~\ref{FracNaI-Final}(b)). This reduction may eventually prove to be too extreme and at present we do not have a better formalism to model dechanneling.

With the simple model of dechanneling we used we could reproduce the channeling fractions computed by the DAMA collaboration (which, however, apply to ions which start their motion close to the middle of a channel and not to the case of WIMP direct detection). For completeness, in Appendix C we also include a model of planar channeling proposed by Matyukhin~\cite{Matyukhin} , which produces larger channeling probabilities, but which we think is not correct.

 The analytical  approach used here can successfully describe qualitative features of the channeling and blocking effects, but should be complemented by data fitting of parameters and by simulations to obtain a good quantitative description too.  Thus our results should in the last instance be checked by using some of the many sophisticated Monte Carlo simulation programs implementing the binary collision approach or mixed approaches.

\begin{acknowledgments}
G.G.   was supported in part by the US Department of Energy Grant
DE-FG03-91ER40662, Task C.  P.G. was  supported  in part by  the NFS
grant PHY-0756962 at the University of Utah. We would like to thank S. Nussinov and F. Avignone for
several important discussions about their work, and  J. U Andersen, D. S.  Gemmell, D. V. Morgan, G. Hobler, and Kai Nordlund for some exchange of information.
We also thank P. Belli for several conversations.
\end{acknowledgments}

\appendix
\addappheadtotoc

\section{Crystal structure and other data for NaI}

\begin{figure}[h]
\begin{center}
  \includegraphics[height=210pt]{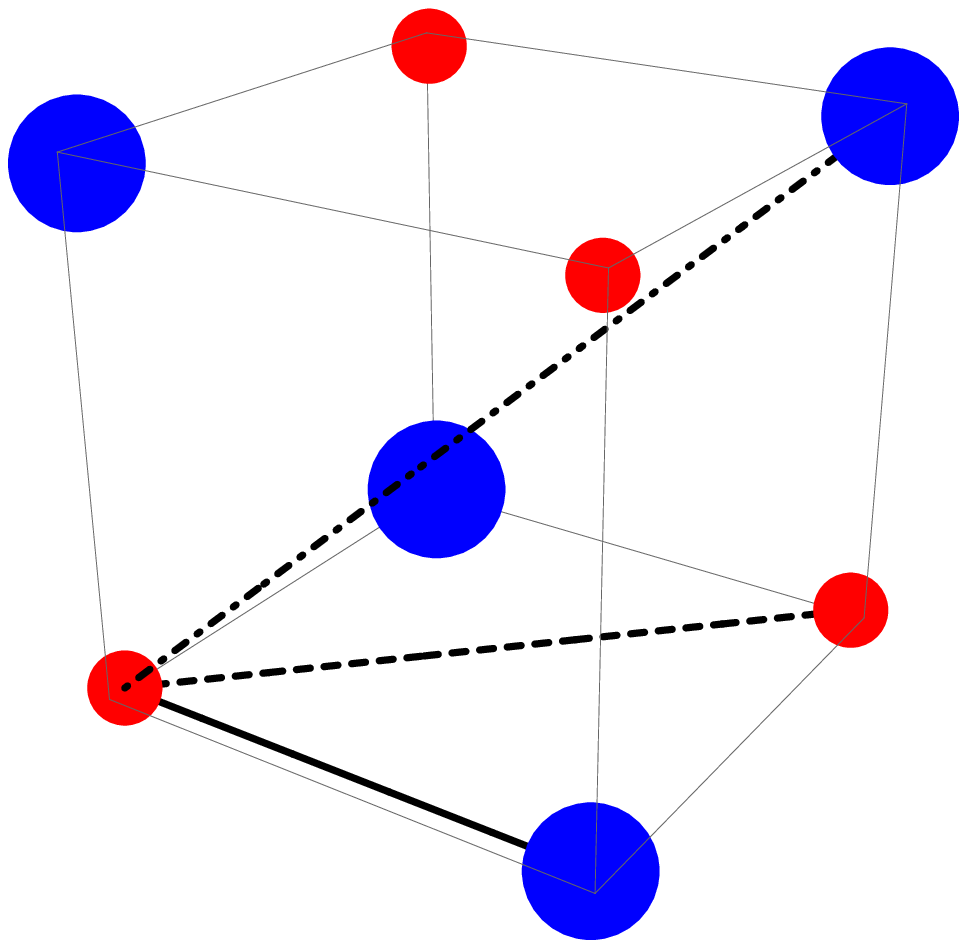}
  \vspace{-0.5cm}\caption{One eights of the NaI crystal unit cell with the red and blue spheres representing Na and I ions respectively. The solid, dashed and dot-dashed lines show the $<$100$>$, $<$110$>$, and $<$111$>$ axes respectively. The \{100\}, \{110\} and \{111\} planes are perpendicular to the respective axes with equal indices.}
  \label{UnitCell}
\end{center}
\end{figure}

NaI is a diatomic compound that has two interpenetrating face-centered cubic (f.c.c.) lattice structures displaced by half of a lattice constant with 8 atoms per unit cell. The lattice constant $a_{\rm lat}$ of a cubic crystal system refers to the constant distance between unit cells of one of the f.c.c. lattices in the crystal, and  for NaI it is $a_{\rm lat}=$0.6473 nm at room temperature (Table 3.4 in Appleton and Foti~\cite{Appleton-Foti:1977}).
 Fig.~\ref{UnitCell} shows one eights of the unit cell of the NaI crystal. The red and blue spheres represent Na and I ions respectively. The shortest distance  between Na and I ions in Fig.~\ref{UnitCell} is half the lattice constant, $a_{\rm lat}/2$.

The atomic mass and atomic number of Na and I are $M_{\rm Na}=22.9$ amu, $M_{\rm I}=126.9$ amu, $Z_{\rm Na}=11$ and $Z_I=53$. When computing $\psi_1$ in Eq.~\ref{psi1} for Na recoils, we take $Z_1=Z_{\rm Na}$ and $Z_2$ equal to an effective atomic number of the row or plane in the channel, which depends on the composition of the row or plane. ``Mixed" channels, for example the rows $<100>$ and $<111>$, or the planes $\{100\}$ and $\{110\}$, contain both Na and I ions in alternation; they have $Z_2=\bar{Z}=(Z_{\rm Na}+Z_{\rm I})/2$. ``Pure" channels, for example the row $<110>$ or the plane $\{111\}$, contain atoms of a single species, only Na or only I; they have $Z_2=Z_{\rm Na}$ or $Z_2=Z_{\rm I}$. Thus, for Na recoils from the row where it originally was, we have
\begin{equation}
\sin^2 \psi_{1}^{\rm Na}=\frac{2Z_{\rm Na}{Z_2}e^2}{E d}.
\end{equation}
Similarly, for I recoils $Z_1=Z_I$,  and for mixed channels  $Z_2=\bar{Z}=(Z_{\rm Na}+Z_{\rm I})/2$ while for pure channels we use $Z_2=Z_{\rm I}$. We have
\begin{equation}
\sin^2 \psi_{1}^{\rm I}=\frac{2Z_{\rm I}{Z_2}e^2}{E d}.
\end{equation}

With respect to  the Thomas-Fermi screening distance,  for Na recoils from a mixed row or plane we use the average
\begin{equation}
 \bar{a}_{\rm Na}=(a_{\rm NaNa}+a_{\rm NaI})/2=0.01149\ {\rm nm},
 \end{equation}
 where $a_{\rm NaNa}=0.4685 (Z_{\rm Na}^{1/2} + Z_{\rm Na}^{1/2})^{-2/3}=0.01327$ nm and $a_{\rm NaI}=0.4685 (Z_{\rm Na}^{1/2} + Z_{\rm I}^{1/2})^{-2/3}=0.009711$ nm  correspond to an Na scattering from an Na and an I lattice atom, respectively. On the other hand, for Na recoils from a pure row or plane we use $a_{\rm NaNa}$ because the row or plane from which the recoiling Na ion was emitted contains  only Na atoms. Similarly, for I recoils from  a mixed row or plane, we use
\begin{equation}
 \bar{a}_{\rm I}=(a_{\rm II}+a_{\rm NaI})/2=0.008784\ {\rm nm},
 \end{equation}
 where $a_{\rm II}=0.4685 (Z_{\rm I}^{1/2} + Z_{\rm I}^{1/2})^{-2/3}=0.007857$ nm  and $a_{\rm NaI}$ correspond to an I ion scattering from an I and an Na lattice atom, respectively. For I recoils from a pure row or plane we use $a_{\rm II}$, since the row or plane the recoiling ion is emitted from is made of I ions only.

Finally, to compute $\psi_a$ for Na recoils
\begin{equation}
\sin\psi_a=\left(\frac{2\pi n Z_{\rm Na} Z_2 e^2 a}{E}\right)^\frac{1}{2},
\end{equation}
where $Z_2=\bar{Z}$ and $a=\bar{a}_{\rm Na}$  or $Z_2=Z_{\rm Na}$ and $a={a}_{\rm NaNa}$ for mixed or pure
rows and planes respectively. For I recoils
\begin{equation}
\sin\psi_a=\left(\frac{2\pi n Z_{\rm I} Z_2 e^2 a}{E_R}\right)^\frac{1}{2},
\end{equation}
where $Z_2=\bar{Z}$ and $a=\bar{a}_{\rm I}$  or $Z_2=Z_{\rm Na}$ and $a={a}_{\rm II}$ for mixed or pure
rows and planes respectively.

We here review the crystallographic notation for directions in the lattice. Once an origin of the coordinate system is fixed on a lattice point $O$, any position vector of a point on the crystal lattice can be written as $\textbf{R}=n_1\textbf{a}+n_2\textbf{b}+n_3\textbf{c}$ with $n_1$, $n_2$, and $n_3$ specific integer numbers. The vectors $\textbf{a}$, $\textbf{b}$, and $\textbf{c}$ are the basis vectors of the crystal lattice, and are three noncoplanar vectors joining the lattice point $O$ to its near neighbors. For the cubic lattice of NaI, the three vectors $\textbf{a}$, $\textbf{b}$, $\textbf{c}$ form a Cartesian frame and their length is $a_{\rm lat}/2$ (they are the sides of the cube in Fig.~\ref{UnitCell}). The integers $n_1$, $n_2$, and $n_3$ can be positive, negative, or zero. The direction of a crystal axis pointing in the direction $\textbf{R}$ is specified by the triplet $[n_1 n_2 n_3]$ written in square brackets, when $n_1$, $n_2$, and $n_3$ are positive or zero. Note that if there is a common factor in the numbers $n_1$, $n_2$, $n_3$, this factor is removed. Moreover, negative integers are denoted with a bar over the number, e.g.\ $-1$ is denoted as $\bar{1}$ and the $-y$ axis is $[0\bar{1}0]$ direction. Fig.~\ref{UnitCell} shows the directions of the $[$100$]$, $[$110$]$ and $[$111$]$ axes.

In a cubic crystal, because of the symmetry of the unit cell, the directions $[100]$, $[010]$, and $[001]$ are equivalent. All directions equivalent to the $[n_1 n_2 n_3]$ direction are denoted by $<$$n_1n_2n_3$$>$ in angular brackets. For example, $<$100$>$ indicates all six directions $[100]$, $[010]$, $[001]$, $[\bar{1}00]$, $[0\bar{1}0]$, and $[00\bar{1}]$. The plane perpendicular to the $[n_1 n_2 n_3]$ axis is denoted by $(n_1 n_2 n_3)$. For example, the plane perpendicular to the $[100]$ axis is denoted by $(100)$, and that perpendicular to $[101]$ by $(101)$. The integers $n_1$, $n_2$, and $n_3$ are called Miller indices.

When the unit cell has cubic symmetry, we can indicate all planes that are equivalent to the plane $(hkl)$ by curly brackets $\{hkl\}$. For example, the indices \{100\} refer to the six planes $(100)$, $(010)$, $(001)$, $(\bar{1}00)$, $(0\bar{1}0)$, and $(00\bar{1})$. The negative sign over a number denotes that the plane cuts the axis on the negative side of the origin.

We will only consider the lower index crystallographic axis and planes. For axial channeling we will consider the $<$100$>$, $<$110$>$ and $<$111$>$ axes and for planar channeling we consider the \{100\},\{110\} and \{111\} planes perpendicular to them.

To compute the interatomic spacing $d$ in axial directions and the interplanar spacing $d_{pch}$ in planar directions, we have to multiply the lattice constant by the following coefficients~\cite{Gemmell:1974ub}:
\begin{itemize}
  \item Axis: $<100>: 1/2$ , $<110>: 1/\sqrt{2}$ , $<111>: \sqrt{3}/2$
  \item Plane: $\{100\}: 1/2$ , $\{110\}: 1/2\sqrt{2}$ , $\{111\}: 1/2\sqrt{3}$
\end{itemize}

For NaI, the Debye temperature is $\Theta=165$ K, and the crystals in the DAMA experiment are at a temperature of 20 $^\circ$C or $293.15$ K.

\section{HEALPix Pixelization}

The Hierarchical Equal Area iso-Latitude Pixelization (HEALPix)~\cite{HEALPix:2005} provides a convenient  way of dividing the surface of a sphere into equal area sectors. An integral over directions can then be performed as a simple Riemann sum. HEALPix has been introduced to pixelize data on a sphere and has been used by cosmic microwave background experiments like WMAP and BOOMERANG.

In HEALPix, the base resolution comprises 12 pixels in three rings: one ring around the north cap, one ring around the south cap, and one ring around the equator. At a higher resolution, each base pixel in each ring is divided into smaller pixels of equal area. The resolution parameter of the grid is $N_{\rm side}$ and it defines the number of divisions along the side of a base-resolution pixel which is needed to obtain a partition with higher resolution. We choose the resolution parameter of the grid to be $N_{\rm side}=50$.

A HEALPix map has $N_{\rm pixel}=12N_{\rm side}^2$ pixels, each with the same area. The angular resolution of the map can be estimated by computing the solid angle covered by each pixel $\Omega=4\pi/N_{\rm pixel}$, and finding the typical diameter of each pixel as if it were small and of circular shape, $\theta_{\rm res}=2\sqrt{\Omega/\pi}$. This gives $\theta_{\rm res} = 2/(\sqrt{3}N_{\rm side})=66.2^\circ/N_{\rm side}$. By choosing $N_{\rm side}=50$, we have 30,000 pixels on the sphere, and a resolution of 1.3 degrees. If the HEALPix is properly aligned with the cubic crystal so that there is a pixel in each $<$100$>$ direction, this resolution should be sufficient for computing our integrals accurately. We have tried different values of $N_{\rm side}$, up to $N_{\rm side}=400$ (for which the resolution is $\theta_{\rm res}=0.166^\circ$), and we found that the value of the channeling fraction already converges within one percent for $N_{\rm side}=20$. Thus, we used $N_{\rm side}=50$ as a safe value in out calculations.

The following algorithm~\cite{HEALPix:2005}  was used to generate a list of unit vectors on a sphere, with each unit vector in the direction of one of the HEALPix pixels. Let $p$ be the pixel index, with $p=0,1,\ldots,N_{\rm pixel}-1$. We start with the definitions:
\begin{equation}
p_{\rm max}=12N_{\rm side}^2 - 1, \qquad p_{\rm max}^{\rm north}=\frac{p_{\rm max} + 1}{2} + 2N_{\rm side} - 1.
\end{equation}
If $p > p_{\rm max}^{\rm north}$, then $q = p_{\rm max} - p$ otherwise $q = p$. Then we define
\begin{equation}
n_{\rm polar} = N_{\rm side} - 1,\qquad
n_{\rm equatorial} = N_{\rm side} + 1,\qquad
p_{\rm min}^{\rm equatorial} = 2 N_{\rm side}(N_{\rm side} - 1).
\end{equation}
We proceed to compute the cylindrical coordinates $z$ and $\varphi$ of the direction of the $p$-th HEALPix pixel. If $q \geq p_{\rm min}^{\rm equatorial}$, the pixel belongs to one of the polar rings and we successively compute
\begin{eqnarray}
h &=& p - p_{\rm min}^{\rm equatorial},~~~~\\
i &=& \left\lfloor \frac{ph}{4 N_{\rm side}}\right\rfloor + N_{\rm side},~~~~\\
j &=& \left[h \bmod{(4 N_{\rm side})}\right] + 1,\\
z &=& \frac{4}{3} - \frac{2i}{3 N_{\rm side}},~~~~\\
s &=& (i - N_{\rm side} + 1) \bmod{2},~~~~\\
\varphi &=& \frac{\pi}{2 N_{\rm side}} \left(j - \frac{s}{2}\right).
\end{eqnarray}
Here $\lfloor x\rfloor$ is the minimum integer less or equal to $x$, and $x \bmod{y}$ is the remainder of the integer division of $x$ by $y$.
If $q < p_{\rm min}^{\rm equatorial}$, the pixel belongs to the equatorial ring and we successively compute
\begin{eqnarray}
h &=& \frac{q + 1}{2},~~~~\\
i &=& \left\lfloor\sqrt{h - \sqrt{\lfloor h\rfloor}}\right\rfloor + 1,~~~~\\
j &=& q + 1 - 2 i (i - 1),\\
\text{If} &&\!\!\!\!\!\!\!\! p > p_{\rm max}^{\rm north} \text{~then~}  j = 4 i -j+1,~~~~\\
z &=& 1 - \frac{i^2}{3 N_{\rm side}^2},~~~~\\
\text{If} &&\!\!\!\!\!\!\!\!  p > p_{\rm max}^{\rm north}  \text{~then~}  z = -z,\\
\varphi &=& \frac{\pi}{2 i} \left(j - \frac{1}{2}\right).~~~~
\end{eqnarray}
Finally, the direction vector of  the $p$-th pixel is given by
\begin{equation}
\hat{\bf n}_p=\left( \sqrt{1 - z^2} \cos{\varphi} , \; \sqrt{1 - z^2} \sin{\varphi} , \; z\right).
\end{equation}

\section{Matyukhin model of planar channeling}

 Matyukhin in Ref~\cite{Matyukhin} uses a different condition for planar channeling than we used previously which consists of Eq.~\ref{U''} written in terms of the planar potential, namely  $U''_p(x) <  8 E/d_p^2 $ (where  $''$ denotes the second derivative with respect to $x$).  We have not been able either to derive this condition for the planar potential or to find the derivation of this condition anywhere. We have not found
the results derived from this condition compared with data either, but we mention the model for completeness.  For  Lindhard's planar potential, the condition used by Matyukhin becomes
\begin{equation}
E > \frac{{d_p}^2}{8} \, \frac{C^2 a E_2}{[x_{\rm min}^2 +C^2 a^2]^{3/2}},
\label{planarU''}
\end{equation}
where $x_{\rm min}$ is the minimum distance of approach  to the plane, $d_p=1/\sqrt{N d_{\rm pch}}$ and
\begin{equation}
E_2=E \psi_a^2=2\pi N d_p Z_1 Z_2 e^2 a.
\end{equation}
 From Eq.~\ref{planarU''},  the smallest possible value of   $x_{\rm min}$  is now
\begin{equation}
x^M_c(E)=C a \sqrt{\left(\frac{{d_p}^2 E_2}{8 C a^2 E}\right)^{2/3} -1}.
\label{xcritM}
\end{equation}
\FIGURE[h]{\epsfig{file=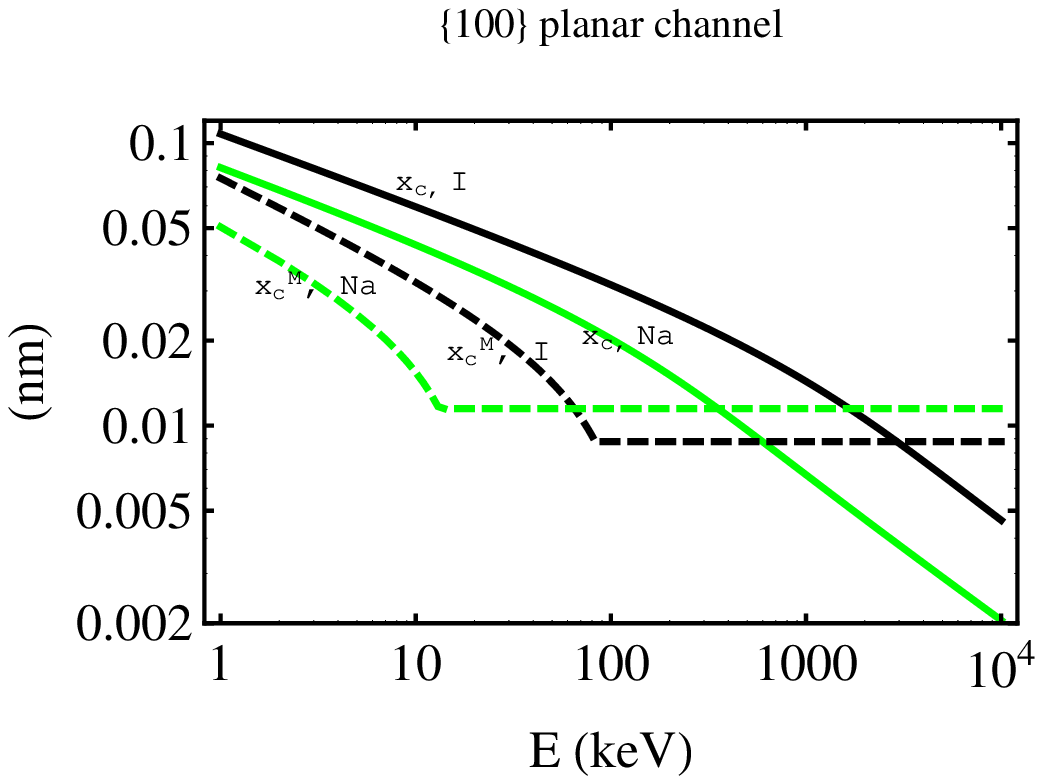,height=165pt}
\epsfig{file=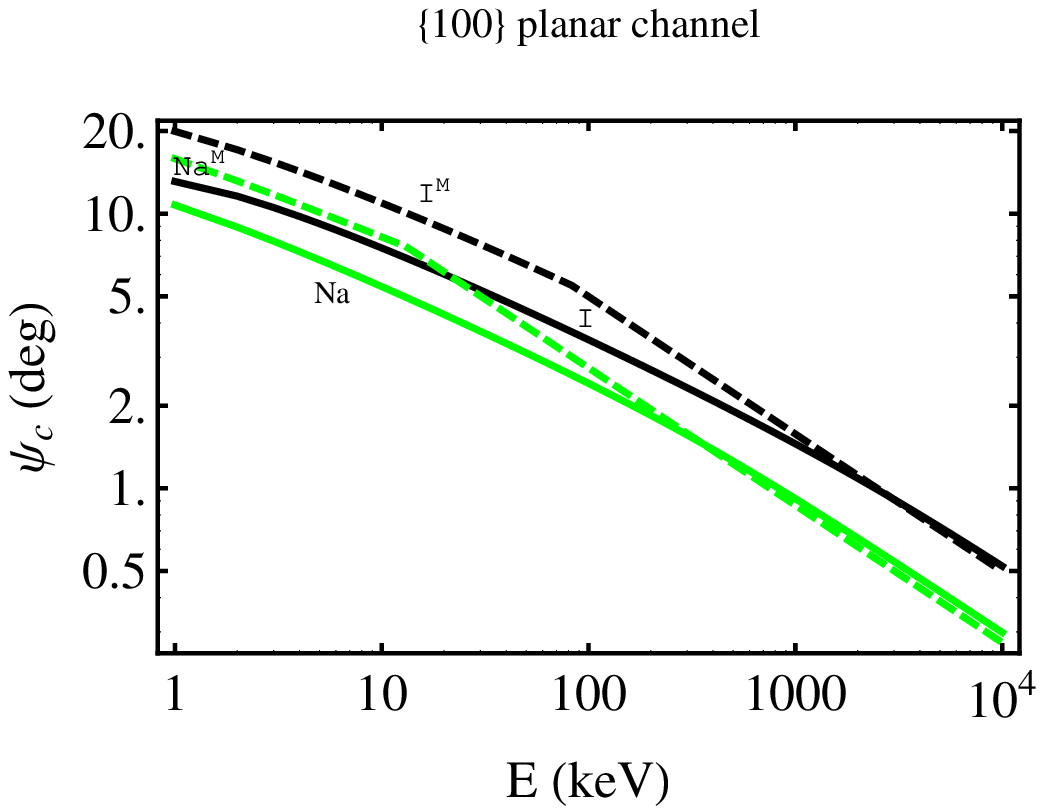,height=165pt}
\epsfig{file=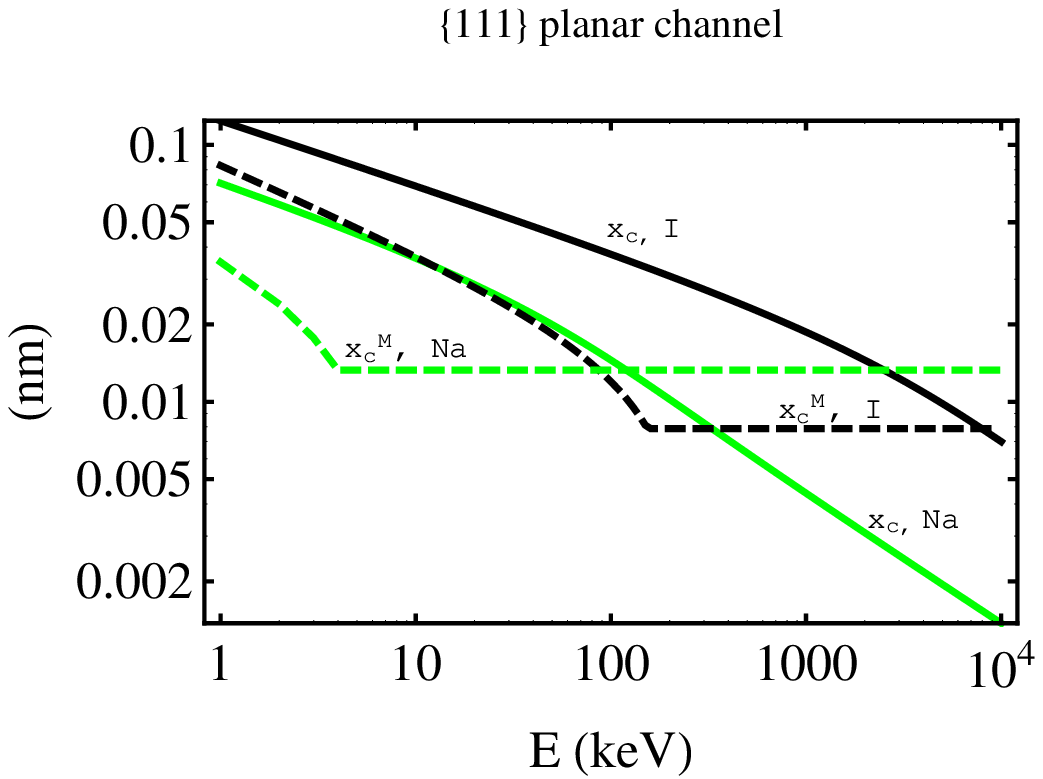,height=165pt}
\epsfig{file=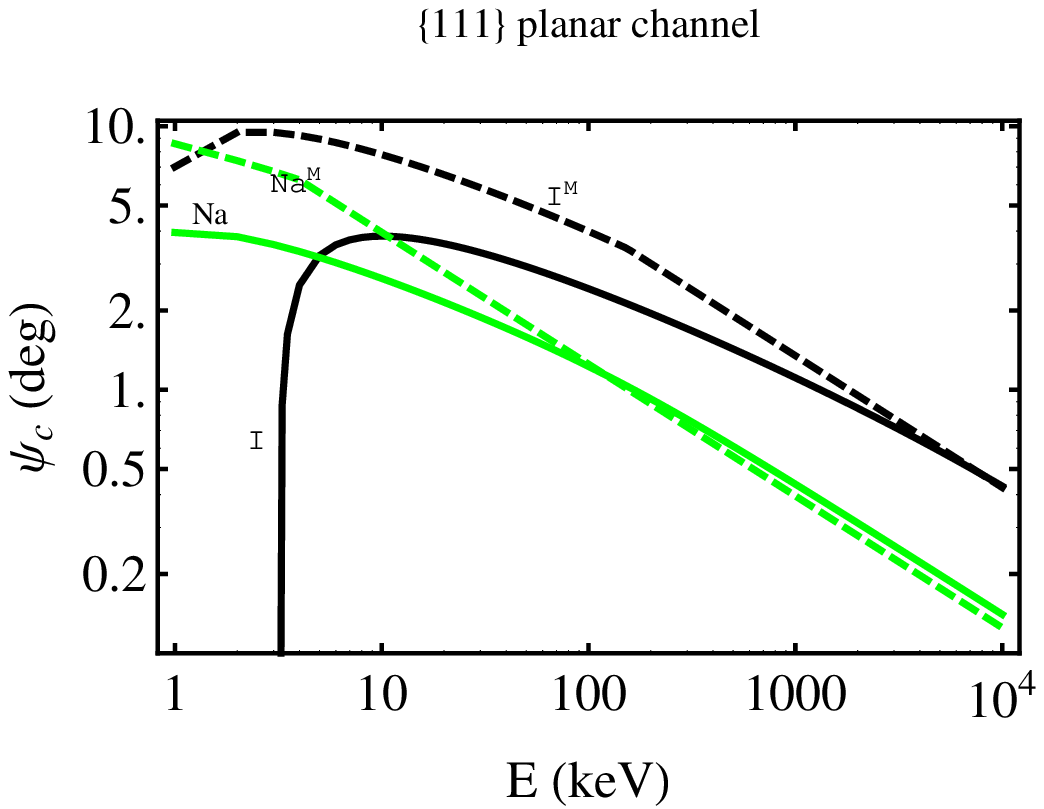,height=165pt}
        \caption{Comparison of \ref{rc-psic-Matyukhin}(a)  (top left) critical distances of approach and \ref{rc-psic-Matyukhin}(b) (top right) critical angles at 20 $^\circ$C with $c_1=c_2=1$ in the \{100\} planar channel predicted by our main model (solid lines, see Figs.~\ref{rc-psic-MV-100}, \ref{rc-psic-MV-111}) and by Matyukhin's (dashed lines). \ref{rc-psic-Matyukhin}(c) (bottom left) and \ref{rc-psic-Matyukhin}(d) (bottom right), same for the \{111\} planar channel. Green/gray lines are for Na and black for I propagating ions.}%
	\label{rc-psic-Matyukhin}}
 For all the energies we consider here ($\sim$ keV and above),  $x^M_c$ is smaller than $d_{\rm pch}/2$, thus $ U_p(d_{\rm pch}/2)$ can be neglected in the definition of the critical planar angle $\psi^p_c$, Eq.~\ref{psicritplanar}.  At low energies $E \leq {d_p}^2 E_2 (8 C a^2)^{-1}$ we have $x^M_c(E) \simeq C a ({d_p}^2 E_2 /8 C a^2 E)^{1/3}=(C^2 \pi Z_1 Z_2 e^2 a^2/4E)^{1/3}$ and
\begin{equation}
\psi_{c}^{pM}(E) \simeq \left(\frac{C^2~a~E_2}{{d_p}~E}\right)^{1/3}=\left(\frac{2~ C^2 \pi Z_1 Z_2 e^2 (N d_{\rm pch})^{3/2} a^2}{E}\right)^{1/3},
\label{psic-M}
\end{equation}
where  $\psi_{c}^{pM}(E) = (6 \pi)^{1/3} \theta_{pl}$  and $\theta_{pl}$ is the critical angle  used by DAMA.

 As $E$ approaches the value ${d_p}^2 E_2 (8 C a^2)^{-1}$, $x^M_c(E)$  approaches zero. At larger energies,  $E > {d_p}^2 E_2 (8 C a^2)^{-1}$, $x^M_c(E)$ in Eq.~\ref{xcritM} becomes imaginary, but  $x^M_c(E)$  is by definition real and positive thus one could take it to be zero. Matyukhin takes in this case  $x^M_c(E)=a$ instead. Either way,  in this  energy range
$\psi_{c}^{pl}(E) \simeq \psi_a$ which is the value given by Lindhard for the ``breakthrough" angle~\cite{Barrett:1971}  necessary to have $E_{\rm perp}= U_p(0)$. We take $x^M_c(E)=a$ wherever the prediction of
Eq.~\ref{xcritM} is smaller than $a$.  Including temperature corrections due to the thermal and zero point energy vibrations of the atoms in the lattice, we have $x_c(T)$ as  in Eq.~\ref{xcofT}.

 The equations of Matyukhin coincide with those presented here if $C=1$ (but, following Lindhard, we take $C=\sqrt{3}$ instead).
\FIGURE{\epsfig{file=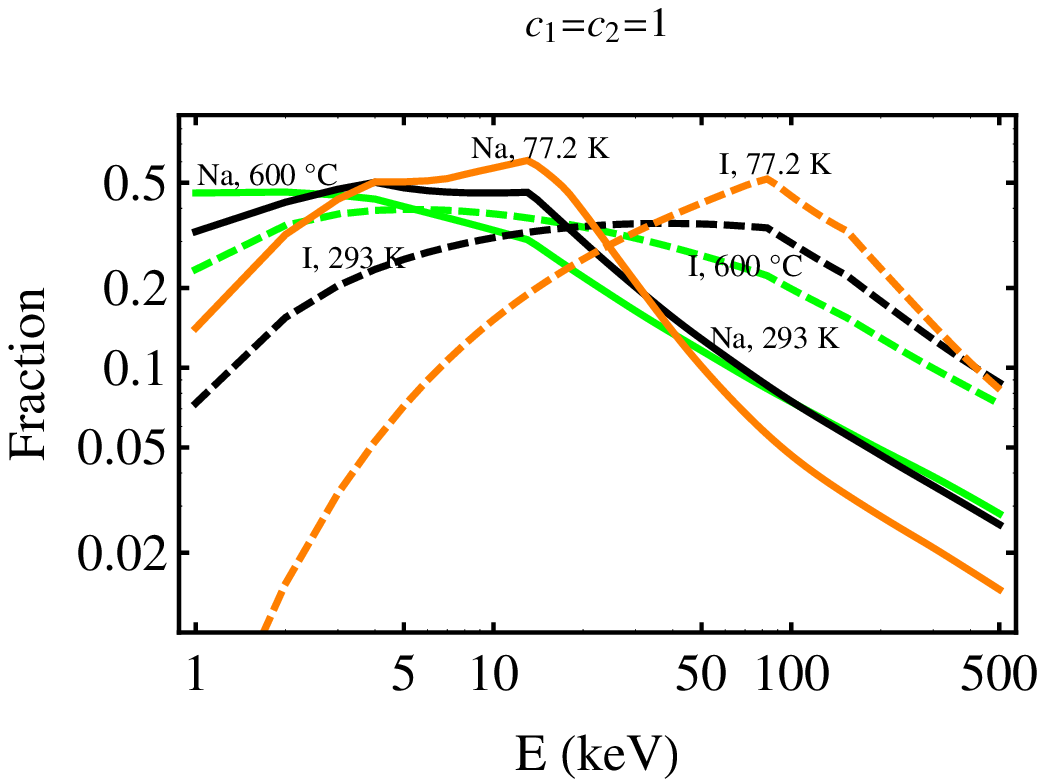,height=163pt}
\epsfig{file=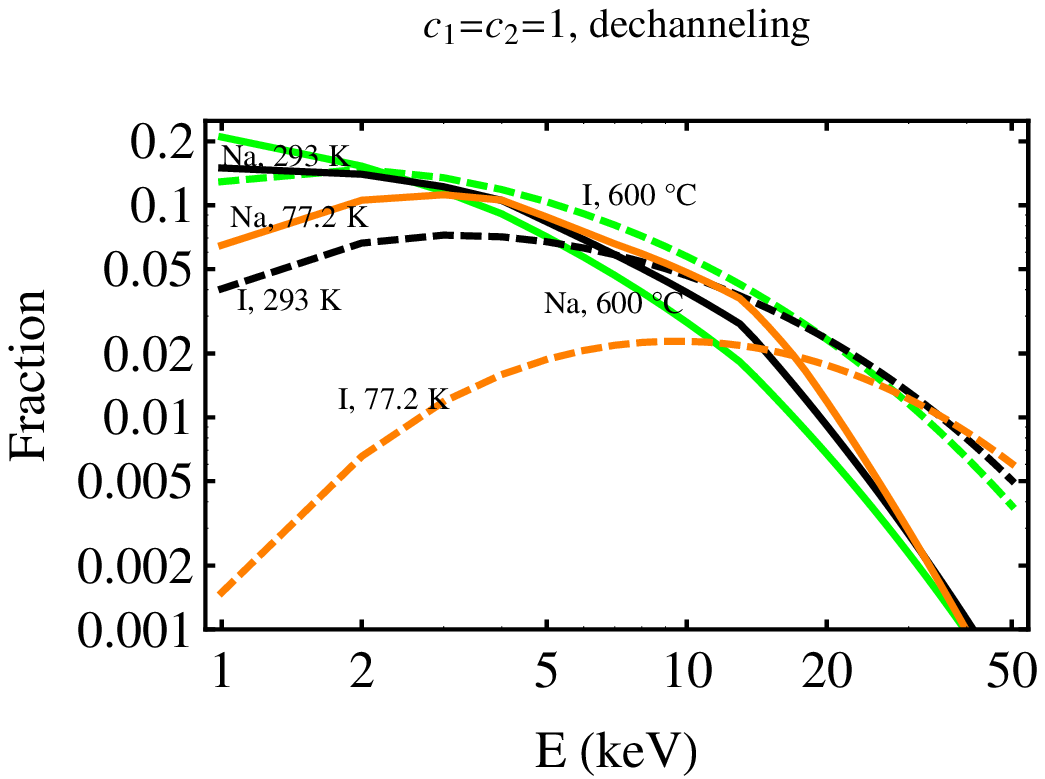,height=163pt}
        \caption{Channeling fractions using Matyukhin's model for the planar channel as a function of the energy of recoiling Na and I ions for T=600 $^\circ$C (green/light gray), 293 K (black) and 77.2 K (orange/dark gray) for  T-corrections included in the lattice  with  $c_1=c_2=1$, (a) without and (b) with dechanneling included as in Eq.~\ref{Pfully-channeled}. The probabilities are larger than in our main method, but we do not trust Matyukhin's approach.}%
	\label{FracNaI-DiffT-c1-M}}
\FIGURE{\epsfig{file=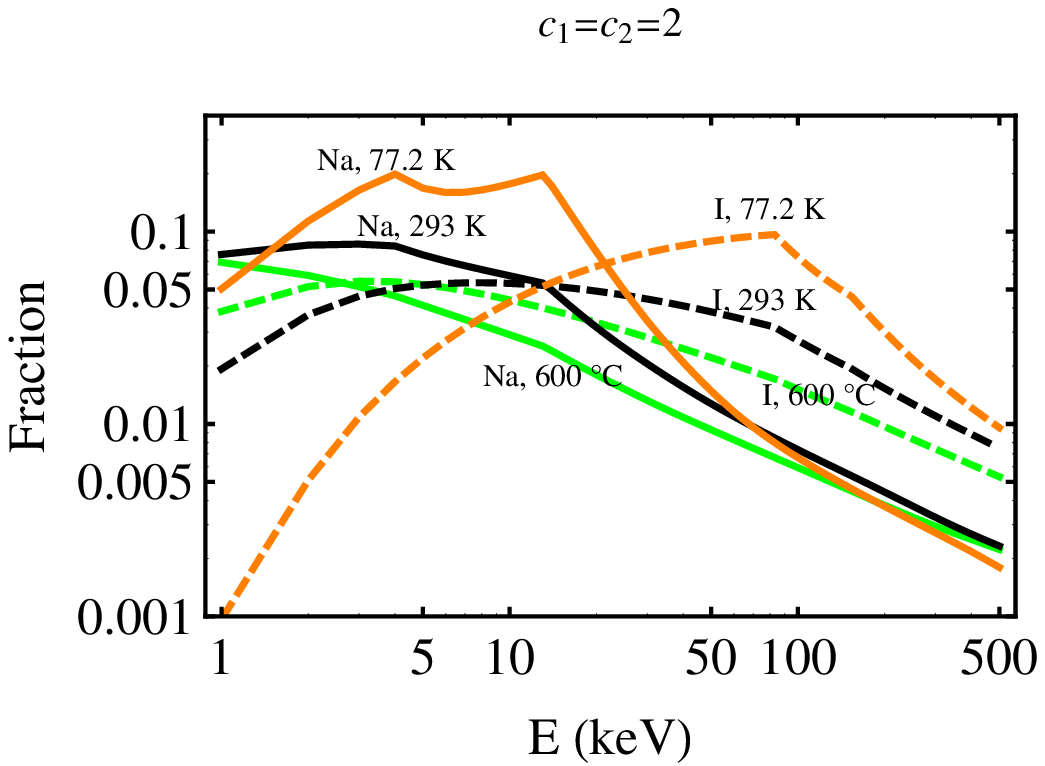,height=160pt}
\epsfig{file=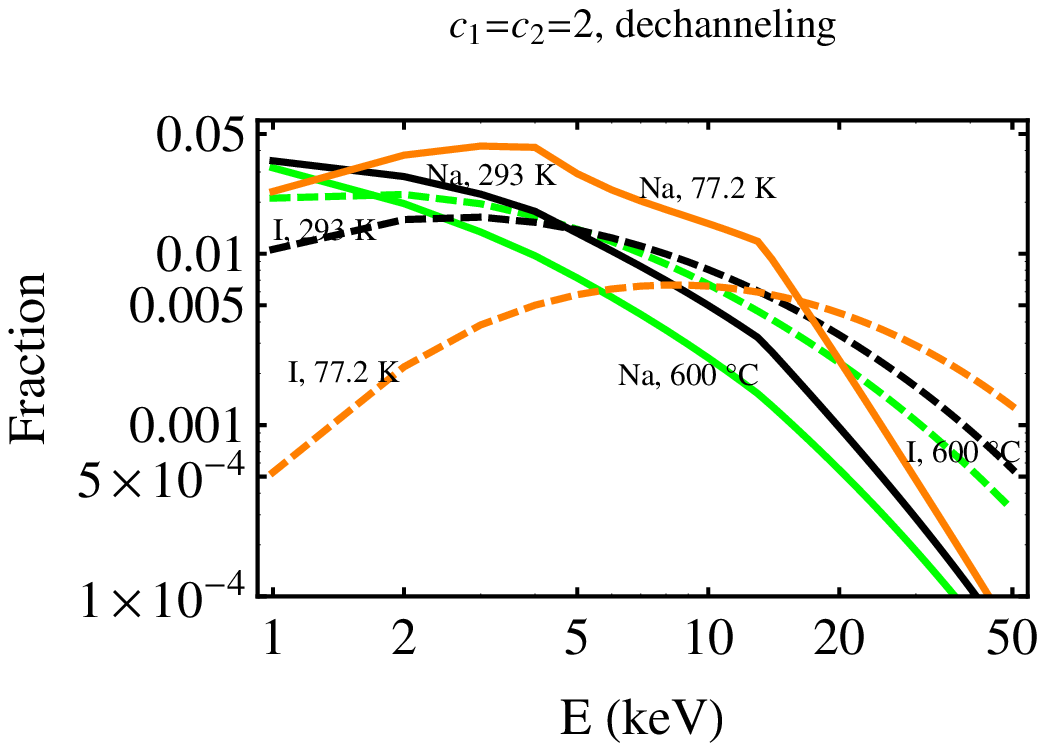,height=160pt}
        \caption{Same as Fig.~\ref{FracNaI-DiffT-c1-M} (for Matyukhin's model) but for $c_1=c_2=2$.}%
	\label{FracNaI-DiffT-c2-M}}

A comparison  of the static critical distances of approach $x_c(E)$ in our method (using Eq.~\ref{ourxcrit} and \ref{psicritplanar}) and in Matyukhin's model, and of the critical angles for $c_1=c_2=1$ in both models is shown in  Figs.~\ref{rc-psic-Matyukhin}(a) and  \ref{rc-psic-Matyukhin}(c) and  in  Figs.~\ref{rc-psic-Matyukhin}(b) and \ref{rc-psic-Matyukhin}(d) respectively for the \{100\} and \{111\} planar channels respectively. The Matyukhin critical distances of approach are smaller
(and thus the critical angles larger) than those in our main method at low energies, which leads to higher channeling fractions, as shown in  Figs.~\ref{FracNaI-DiffT-c1-M} and \ref{FracNaI-DiffT-c2-M} for $c_1=c_2=1$ and for $c_1=c_2=2$ respectively. In these figures the left panels are without and the right panels with dechanneling  as in Eq.~\ref{Pfully-channeled} included. The channeling fractions using Matyukhin's model are much higher than the fractions we obtain with our method, but, as explained above, we do not trust Matyukhin's model. The critical distances in Matyukhin's model (see Figs.~\ref{rc-psic-Matyukhin}(a) and  \ref{rc-psic-Matyukhin}(c)) have a discontinuous slope at the energy where they become constant and this shows in the channeling fraction curves also as discontinuities in slope (at the values of $E$ at which different important channels  have a sharp change in $x_c^M$).

\section{Probability of correlated channels}

In Eq.~\ref{Prob} we treat channeling along different channels as independent events. Here we prove that this procedure is adequate for our purpose of providing upper bounds to the channeling fractions.

For an axial channel when $\phi<\psi_c$ (otherwise $\chi_{\rm axial}=0$), the integration region in Eq.~\ref{chiaxial} is the exterior of an infinitely long cylinder of radius $r_{i, {\rm min}}(E, \phi)$ and axis coincident with the channel axis. Similarly, for a planar channel when $\phi<\psi_c^p$ (otherwise $\chi_{\rm planar}=0$), the integration region in Eq.~\ref{chiplanar} is the exterior of an infinite slab of half-thickness $x_{i, {\rm min}}(E, \phi)$. Let us consider the complements of the integration regions, i.e. the regions excluded in the integrals. These are the regions interior to a cylinder (for an axial channel) or a slab (for a planar channel).

Note that only channels making angles with the direction $\hat{\bf q}$ (of the initial momentum) smaller than their respective critical angles contribute to the union of integration regions. Therefore the problem of combining channels arises only when a recoil direction belongs to more than one channel, and this happens if the channels overlap for some directions of $\hat{\bf q}$. For the cases we consider (cubic lattices), only axial channels overlap with a subset of planar channels, or two or more planar channels overlap with each other. Notice that two different planar channels crossing at an angle, overlap in a parallelepiped of very long length on one side, thus one can define an inscribed cylinder within the parallelepiped, whose diameter is equal to the smallest of the two widths of both planar channels.

We can obtain an upper limit to the channeling probability of overlapping channels by replacing the intersection of the complements of the integration regions with the inscribed circle of radius $r_{\rm MIN}$  equal to the minimum of the $r_{i, {\rm min}}$ or $x_{i, {\rm min}}$ among the overlapping channels. Then an upper bound to the probability $\chi_{\rm rec}(E, \hat{\bf q})$ in Eq.~\ref{probrec} is
\begin{equation}
\chi_{\rm rec}(E, \hat{\bf q}) \leq \int_{r_{\rm MIN}}^{\infty}{dr g(r)}=\exp{(-r_{\rm MIN}^2/2u_1^2)}.
\label{Maxprob}
\end{equation}
When only one channel is open (i.e. $\phi<\psi_c$ for only one channel), we still use Eq.~\ref{chiaxial} or \ref{chiplanar} for the channeling probabilities.

Fig.~\ref{FracNaI-Max} shows the comparison of this upper limit with the fractions we computed using Eq.~\ref{Prob}. The two are practically indistinguishable. This proves that the method we used (see Section 5.3) is adequate for our purpose of providing upper bounds to the channeling fractions.
\FIGURE{\epsfig{file=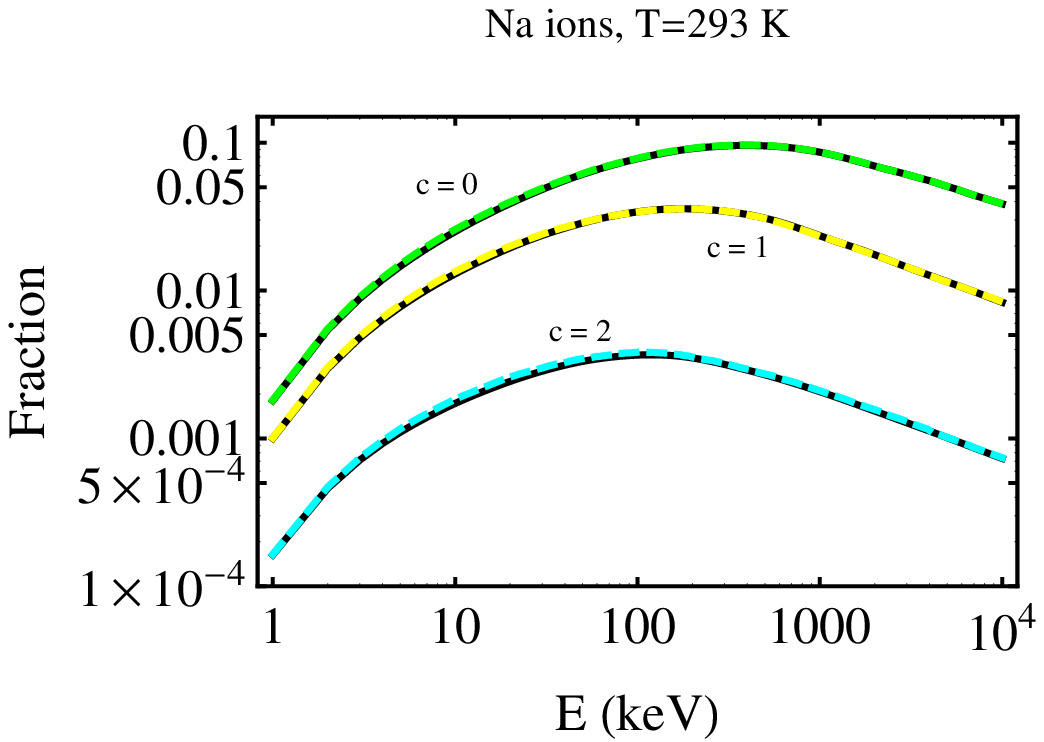,height=150pt}
\epsfig{file=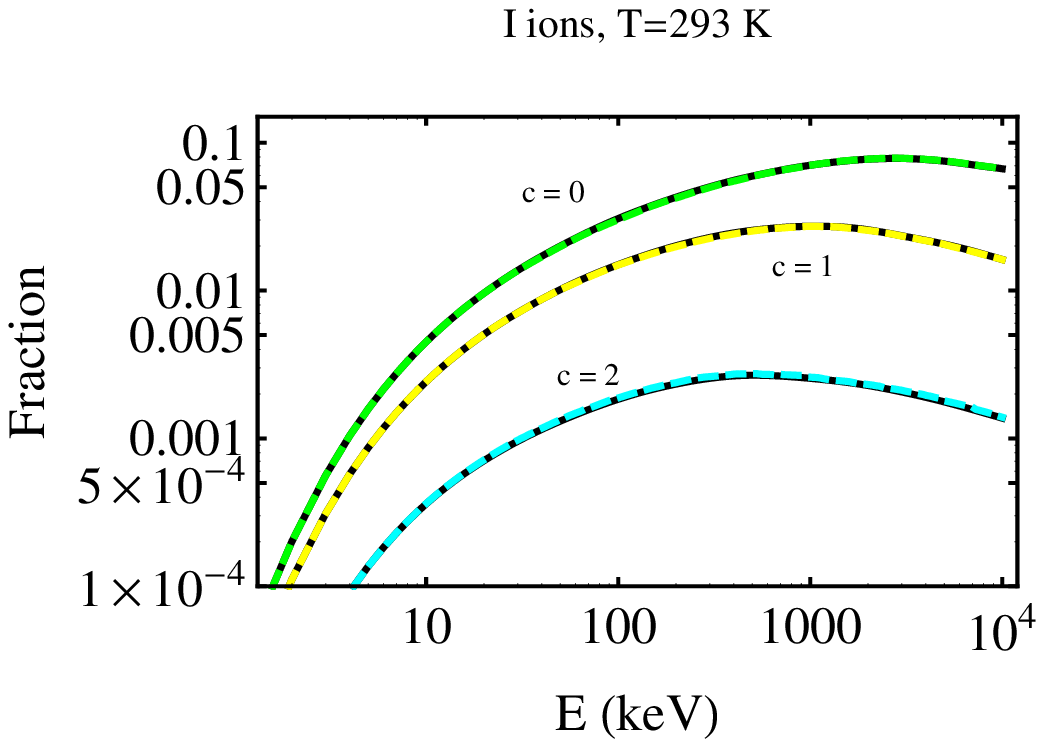,height=150pt}
        \caption{Maximum channeling fractions using Eq.~\ref{Maxprob} for $c_1=c_2=c$ and $c=0$ (dashed green), $c=1$ (dashed yellow) and $c=2$ (dashed cyan) compared with the results of our method of Section 5.3 (solid black lines) for the same models for (a) Na ions and (b) I ions propagating in NaI crystal at $T=293$ K. Notice that the lines overlap.}%
	\label{FracNaI-Max}}


\begin{thebibliography}{99}


\bibitem{Savage:2010}
C.~Savage, G.~Gelmini, P.~Gondolo and K.~Freese,
  arXiv:1006.0972v2 [astro-ph.CO].


\bibitem{Avignone:2008cw}
  F.~T.~Avignone, R.~J.~Creswick and S.~Nussinov,
  arXiv:0807.3758 [hep-ph];
 R.~J.~Creswick, S.~Nussinov and F.~T.~Avignone,
  arXiv:1007.0214v2 [astro-ph.IM].


\bibitem{rol} P.~K.~Rol {\it et al.}
``Proc. Fourth Int.  Conf on Ionization Phenomena in Gases"
ed. by N.R. Nilsson, North Holland, p.257 (1960).

\bibitem{robinson} M. T. Robinson
Appl.\ Phys.\ Lett. {\bf 1}, 49 (1962).


\bibitem{stark} J. Stark,
Phys. Z. {\bf 13}, 913 (1912).

\bibitem{Lindhard:1965} J.~Lindhard,
Kongel.\ Dan.\ Vidensk.\ Selsk.,\ Mat.-Fys.\ Medd.\  {\bf 34}, No. 14 (1965).

\bibitem{Gemmell:1974ub}
  D.~S.~Gemmell,
  Rev.\ Mod.\ Phys.\  {\bf 46}, 129 (1974).


\bibitem{recent-chann}  H. Erramli and G. Blondiaux,
Appl. Radiat. Isot. {\bf 46}, 413 (1995);
J. U Andersen {\it et al.},
Phys.\ Rev.\  C {\bf 78}, 064609 (2008).

\bibitem{cohen} C. Cohen and D. Dauvergne
Nucl. Instrum. Methods Phys. Res.  {\bf B 225}, 40 (2004).

\bibitem{Altman}  R. Altman, H. B. Dietrich, R. B. Murray, and T. J. Rock,
 Phys.\ Rev.\  B {\bf 7}, 1743 (1973).


\bibitem{Drobyshevski:2007zj}
  E.~M.~Drobyshevski,
  Mod.\ Phys.\ Lett.\  A {\bf 23},  3077 (2008)
  [arXiv:0706.3095 [physics.ins-det]].

\bibitem{Bernabei:2007hw}
  R.~Bernabei {\it et al.},
  Eur.\ Phys.\ J.\  C {\bf 53}, 205 (2008)
  [arXiv:0710.0288 [astro-ph]].

\bibitem{implantation} L. Rubin and J. Poate,
The Industrial Physicist, June/July 2003, p.12-15.

\bibitem{Barrett:1971} J. H. Barrett, Phys.\ Rev.\ B{\bf 3}, 1527 (1971).

\bibitem{Monte-Carlo-programs}
MDRANGE, http://beam.acclab.helsinki.fi/knordlun/mdh/mdh\_program.html;
SRIM, http: //www.srim.org/;
TRIM, J. F. Ziegler, "Ion Implantation Technology", Ion Implantation Technology Co. (1996);
MARLOWE and UT-MARLOWE, Y. Chen {\it et al.}, IEEE Trans. Electron Devices, vol. 49, no. 9, 1519 (2002);
Crystal-TRIM, http://www.fzd.de/pls/rois/;
REED, K.~M.~Beardmore and N Gronbech-Jensen, Phys. Rev. B {\bf 60}, 12610 (1999).

\bibitem{SARIC}
V. Bykov {\it et al.}
Nucl. Instrum. Methods Phys. Research (NIM){\bf B 114}, 371 (1996).

\bibitem{MDRANGE}
 K. Nordlund,
 Comput. Mater. Sci. {\bf 3}, 448 (1995).

\bibitem{Hobler}
G. Hobler,
Radiation effects and defects in solids {\bf 139}, 21 (1996);
G. Hobler,
Nucl. Instrum. Methods Phys. Research (NIM){\bf B 115}, 323 (1996).


 \bibitem{cho} K. Cho {\it et al},
 Phys. Research {\bf B}7/8 , 265 (1985).

 \bibitem{andreen} C. J. Andreen and R.L. Hines,
 Phys.\ Rev. {\bf 159}, 285 (1967).

 \bibitem{bergstrom}
 I. Bergstrom {\it et al.},
 Can. J. Phys. {\bf 46}, 2679 (1968).

\bibitem{vanwijngaarden}
A. van Wijngaarten, B. Miremadi, N.M.A. Finney and J.N. Bradford,
 Phys.\ Rev.{\bf 185}, 490 (1969).

\bibitem{lui}
K. M. Lui, Y. Kim, W. M. Lau and J. W. Rabalais,
Jour. Applied Phys {\bf 86}, 5256 (1999).

\bibitem{hogg}
S.M. Hogg, B. Pipeleers, A. Vantomme and M. Swart,
Applied Phys. Lett. {\bf 80}, 4363 (2002).


\bibitem{fang}
Z. L. Fang, W. M. Lau and J. W. Rabalais,
Surface Science {\bf 581}, 1 (2005).

\bibitem{Andersen:1967} J. U. Andersen,
Kongel.\ Dan.\ Vidensk.\ Selsk.,\ Mat.-Fys.\ Medd.\  {\bf 36}, No. 7 (1967).

\bibitem{Morgan-VanVliet}
D. V. Morgan and D. Van Vliet,
Can. J. Phys. {\bf 46}, 503 (1963);
D. V. Morgan and D. Van Vliet,
Radiat. Effects and Defects in Solids {\bf 8}, 51 (1971).

\bibitem{VanVliet}
D. van Vliet in ``Channeling", ed. by  D. V. Morgan (Wiley, London), 37 (1973).

\bibitem{Andersen-Feldman}
J.U. Andersen and L.C. Feldman,
Phys.\ Rev.\ B {\bf 1}, 2063 (1970).

\bibitem{Komaki:1970} K. Komaki and F. Fujimoto,
Phys. Stat.  Sol. (a) {\bf 2},  875 (1970).

\bibitem{Dearnaley:1973}  Chapter 2 of ''Ion Implantation'', by  Geoffrey Dearnaley,
Amsterdam, North-Holland Pub. Co.; New York, American Elsevier, 1973.

\bibitem{Appleton-Foti:1977}
B. R. Appleton and G. Foti, ``Channeling"  in {\it Ion Beam Handbook for Material Analysis}, edited by J. W. Mayer and E. Rimini (Academic, New York), p. 67 (1977).

\bibitem{ion-book}  Chapter 4 of ``Ion-induced electron emission from crystalline solids'', by Hiroshi Kudo,
Springer Tracts in Modern Physics,  2002.

\bibitem{Rozhkov}
V. V. Rozhkov and S. V. DyulÕdya, PisÕma Zh. Tekh. Fiz.
{\bf 10}, 1181 (1984) [Sov. Tech. Phys. Lett. {\bf 10}, 499 (1984)].

\bibitem{Matyukhin}
S.  I. Matyukhin,
Technical Physics {\bf 53}, 1578 (2008).

\bibitem{Lonsdale-1948}
K. Lonsdale, Acta Cryst. {\bf 1}, 142 (1948);
``X-Ray Diffraction" by B. E. Warren, Addison Wesley, p. 61 (1969).

\bibitem{Sharko-Botaki-1971}   A. V. Sharko and A. A. Botaki,
Sov. Phys. Jour. {\bf14}, 330 (1971).

\bibitem{Sharko-BotakiVar-1971}  A. V. Sharko and A. A. Botaki,
Sov. Phys. Jour. {\bf14}, 765 (1971).

\bibitem{Neelakanda-2008}  N. Neelakanda Pillai, C.K. Mahadevan,
Physica B {\bf 403}, 2168  (2008).

\bibitem{Geeta-1998} P. Geeta Krishna, K. G. Subhadra  and D. B. Sirdeshmuk,
Acta Cryst.A{\bf54}, 253 (1998).


\bibitem{Hewat-1972} A. W.  Hewat,
J. Phys. C: Solid State Phys. {\bf 5}, 1309 (1972);
M. Inagaki, M. Toyoda, M. Sakai,
Jour. of Materials Sci. {\bf 22}, 3459  (1987).

\bibitem{HEALPix:2005} K. M. G\'{o}rski {\it et al.},
 ApJ {\bf 622}, 759 (2005).

\bibitem{Lindhard-Scharff} J. Lindhard and M. Scharff,
Phys.\ Rev.\ {\bf 124}, 128 (1961).


\bibitem{Komaki-et-al-1971}  K. Komaki {\it el al.},
Phys. Stat.  Sol. (a) {\bf 4}, 495 (1971).

\bibitem{NGG-WIMPwind} N. Bozorgnia, G. Gelmini and P. Gondolo,
 ``Channeling in direct dark matter detection IV: daily modulation of the WIMP signal",
 in preparation.

 \bibitem{NGG-SiandGe} N. Bozorgnia, G. Gelmini and P. Gondolo,
 arXiv:1008.3676 [astro-ph.CO].




\end{thebibliography}
\end{document}